\documentclass[preprint,showpacs,preprintnumbers,amsmath,amssymb]{revtex4}

\usepackage{graphicx}
\usepackage{dcolumn}
\usepackage{bm}
\usepackage{epsfig}
\usepackage{longtable}

\font\eightit=cmti8

\def\r#1{\ignorespaces $^{#1}$}

\def\TOTXSECBR{  240 \pm 1 (stat) ^{+21}_{-19} (syst)}
\def\TOTXSEC{  4.08 \pm 0.02 (stat)^{+0.36}_{-0.33} (syst)}

\def\XSECGTFIVEETA{  16.3 \pm 0.1 (stat)^{+1.4}_{-1.3} (syst)}
\def\BXSECBR{  19.4 \pm 0.3 (stat)^{+2.1}_{-1.9} (syst)}

\def\BXSECBRFIVEGEV{ 3.06 \pm 0.04 (stat) \pm 0.22 (syst)}

\def\BXSECBRFIVEETA{ 2.75 \pm 0.04 (stat) \pm 0.20 (syst)}

\def\BXSEC{  0.330 \pm 0.005 (stat)  ^{+0.036}_{-0.033} (syst)}

\def\TOTXSECB{  17.6 \pm 0.4 (stat)^{+2.5}_{-2.3} (syst)}

\def\PROMPTXSEC{2.86 \pm 0.01(stat)^{+0.34}_{-0.45} (syst)}

\begin{document}


\preprint{\vbox{\hbox{FERMILAB-PUB-04-440-E\hfill}
                \hbox{Submitted to PRD, December 23, 2004\hfill}}}

\title{Measurement of the \boldmath{$J/\psi$} Meson 
and \boldmath{$b$}-Hadron Production Cross Sections 
in \boldmath{$p\bar{p}$} Collisions at \boldmath{$\sqrt{s}=1960$}\,GeV}

\author{
\begin{sloppypar}
\noindent 
D.~Acosta,\r {16} J.~Adelman,\r {12} T.~Affolder,\r 9 T.~Akimoto,\r {54}
M.G.~Albrow,\r {15} D.~Ambrose,\r {43} S.~Amerio,\r {42}  
D.~Amidei,\r {33} A.~Anastassov,\r {50} K.~Anikeev,\r {15}
% ...
\end{sloppypar}
\vskip .026in
\begin{center}
(CDF Collaboration)
\end{center}
\vskip 0.026in
\begin{center}
\r 1  {\eightit Institute of Physics, Academia Sinica, Taipei, Taiwan 11529, 
Republic of China} \\
\r 2  {\eightit Argonne National Laboratory, Argonne, Illinois 60439} \\
\r 3  {\eightit Institut de Fisica d'Altes Energies, Universitat Autonoma
de Barcelona, E-08193, Bellaterra (Barcelona), Spain} \\
\r 4  {\eightit Istituto Nazionale di Fisica Nucleare, University of Bologna,
I-40127 Bologna, Italy} \\
\r 5  {\eightit Brandeis University, Waltham, Massachusetts 02254} \\
\r 6  {\eightit University of California at Davis, Davis, California  95616} \\
\r 7  {\eightit University of California at Los Angeles, Los 
Angeles, California  90024} \\
\r 8  {\eightit University of California at San Diego, La Jolla, California  92093} \\ 
\r 9  {\eightit University of California at Santa Barbara, Santa Barbara, California 
93106} \\ 
\r {10} {\eightit Instituto de Fisica de Cantabria, CSIC-University of Cantabria, 
39005 Santander, Spain} \\
\r {11} {\eightit Carnegie Mellon University, Pittsburgh, PA  15213} \\
\r {12} {\eightit Enrico Fermi Institute, University of Chicago, Chicago, 
Illinois 60637} \\
\r {13}  {\eightit Joint Institute for Nuclear Research, RU-141980 Dubna, Russia}
\\
\r {14} {\eightit Duke University, Durham, North Carolina  27708} \\
\r {15} {\eightit Fermi National Accelerator Laboratory, Batavia, Illinois 
60510} \\
\r {16} {\eightit University of Florida, Gainesville, Florida  32611} \\
\r {17} {\eightit Laboratori Nazionali di Frascati, Istituto Nazionale di Fisica
               Nucleare, I-00044 Frascati, Italy} \\
\r {18} {\eightit University of Geneva, CH-1211 Geneva 4, Switzerland} \\
\r {19} {\eightit Glasgow University, Glasgow G12 8QQ, United Kingdom}\\
\r {20} {\eightit Harvard University, Cambridge, Massachusetts 02138} \\
\r {21} {\eightit The Helsinki Group: Helsinki Institute of Physics; and Division of
High Energy Physics, Department of Physical Sciences, University of Helsinki, FIN-00044, Helsinki, Finland}\\
\r {22} {\eightit Hiroshima University, Higashi-Hiroshima 724, Japan} \\
\r {23} {\eightit University of Illinois, Urbana, Illinois 61801} \\
\r {24} {\eightit The Johns Hopkins University, Baltimore, Maryland 21218} \\
\r {25} {\eightit Institut f\"{u}r Experimentelle Kernphysik, 
Universit\"{a}t Karlsruhe, 76128 Karlsruhe, Germany} \\
\r {26} {\eightit High Energy Accelerator Research Organization (KEK), Tsukuba, 
Ibaraki 305, Japan} \\
\r {27} {\eightit Center for High Energy Physics: Kyungpook National
University, Taegu 702-701; Seoul National University, Seoul 151-742; and
SungKyunKwan University, Suwon 440-746; Korea} \\
\r {28} {\eightit Ernest Orlando Lawrence Berkeley National Laboratory, 
Berkeley, California 94720} \\
\r {29} {\eightit University of Liverpool, Liverpool L69 7ZE, United Kingdom} \\
\r {30} {\eightit University College London, London WC1E 6BT, United Kingdom} \\
\r {31} {\eightit Massachusetts Institute of Technology, Cambridge,
Massachusetts  02139} \\   
\r {32} {\eightit Institute of Particle Physics: McGill University,
Montr\'{e}al, Canada H3A~2T8; and University of Toronto, Toronto, Canada
M5S~1A7} \\
\r {33} {\eightit University of Michigan, Ann Arbor, Michigan 48109} \\
\r {34} {\eightit Michigan State University, East Lansing, Michigan  48824} \\
\r {35} {\eightit Institution for Theoretical and Experimental Physics, ITEP,
Moscow 117259, Russia} \\
\r {36} {\eightit University of New Mexico, Albuquerque, New Mexico 87131} \\
\r {37} {\eightit Northwestern University, Evanston, Illinois  60208} \\
\r {38} {\eightit The Ohio State University, Columbus, Ohio  43210} \\  
\r {39} {\eightit Okayama University, Okayama 700-8530, Japan}\\  
\r {40} {\eightit Osaka City University, Osaka 588, Japan} \\
\r {41} {\eightit University of Oxford, Oxford OX1 3RH, United Kingdom} \\
\r {42} {\eightit University of Padova, Istituto Nazionale di Fisica 
          Nucleare, Sezione di Padova-Trento, I-35131 Padova, Italy} \\
\r {43} {\eightit University of Pennsylvania, Philadelphia, 
        Pennsylvania 19104} \\   
\r {44} {\eightit Istituto Nazionale di Fisica Nucleare, University and Scuola
               Normale Superiore of Pisa, I-56100 Pisa, Italy} \\
\r {45} {\eightit University of Pittsburgh, Pittsburgh, Pennsylvania 15260} \\
\r {46} {\eightit Purdue University, West Lafayette, Indiana 47907} \\
\r {47} {\eightit University of Rochester, Rochester, New York 14627} \\
\r {48} {\eightit The Rockefeller University, New York, New York 10021} \\
\r {49} {\eightit Istituto Nazionale di Fisica Nucleare, Sezione di Roma 1,
University di Roma ``La Sapienza," I-00185 Roma, Italy}\\
\r {50} {\eightit Rutgers University, Piscataway, New Jersey 08855} \\
\r {51} {\eightit Texas A\&M University, College Station, Texas 77843} \\
\r {52} {\eightit Texas Tech University, Lubbock, Texas 79409} \\
\r {53} {\eightit Istituto Nazionale di Fisica Nucleare, University of Trieste/\
Udine, Italy} \\
\r {54} {\eightit University of Tsukuba, Tsukuba, Ibaraki 305, Japan} \\
\r {55} {\eightit Tufts University, Medford, Massachusetts 02155} \\
\r {56} {\eightit Waseda University, Tokyo 169, Japan} \\
\r {57} {\eightit Wayne State University, Detroit, Michigan  48201} \\
\r {58} {\eightit University of Wisconsin, Madison, Wisconsin 53706} \\
\r {59} {\eightit Yale University, New Haven, Connecticut 06520} \\
\end{center}
\vskip 1.0in
}

\noaffiliation


\begin{abstract}
We present a new measurement of the inclusive and differential
production cross sections of $J/\psi$ mesons and $b$-hadrons in
proton-antiproton collisions at $\sqrt{s}=1960$\,GeV.  The data
correspond to an integrated luminosity of $39.7$ pb$^{-1}$ collected
by the CDF Run II detector.  We find the integrated cross section for
inclusive $J/\psi$ production for all transverse momenta from 0 to 
20\,GeV/$c$ in the rapidity range $|y|<0.6$ to 
be $\TOTXSEC \ \mu {\rm b}$.  We 
separate the fraction of $J/\psi$ events from the decay of the
long-lived $b$-hadrons using the lifetime distribution in all events with 
$p_T(J/\psi) > 1.25$\,GeV/$c$.
We find the total cross section for $b$-hadrons, including both hadrons and  
anti-hadrons, decaying to $J/\psi$ with transverse momenta greater
than $1.25$\,GeV/$c$ in the rapidity range $|y(J/\psi)|<0.6$, 
is $\BXSEC~\mu{\rm b}$.
Using a Monte Carlo simulation of
the decay kinematics of $b$-hadrons to all final states containing a
$J/\psi$, we extract the first measurement of the total single
$b$-hadron cross section down to zero transverse momentum at
$\sqrt{s}=1960$\,GeV. We find the total 
single $b$-hadron cross section
integrated over all transverse momenta 
for $b$-hadrons in the rapidity
range $|y|<0.6$ to be $\TOTXSECB \
\mu{\rm b}$. 

\end{abstract}

\baselineskip12pt   

\maketitle
\baselineskip24pt	
\section{Introduction}


The production of both charmonium mesons and bottom-flavored
hadrons (referred to as $b$-hadrons or $H_b$ in this paper) 
in proton-antiproton colliders has sustained continued interest over the 
last several years. 
There are three major sources of the 
$J/\psi$ mesons: directly produced $J/\psi$, prompt decays of heavier 
charmonium states such as $^3P_1$ state $\chi_{\rm c1}$ and $^3P_2$ state 
$\chi_{c2}$, 
and decays of $b$-hadrons.  Early hadroproduction models of
quarkonium states could not describe the cross section of directly
produced $J/\psi$ mesons. These models under-predicted
the measurements by a factor of approximately 50, and did not
adequately describe the cross-section shape~\cite{RUNIPSI}. With the 
advent of the effective field theory, 
nonrelativistic QCD (NRQCD)~\cite{NRQCD}, better theoretical
descriptions of quarkonium production became possible. Within the
NRQCD factorization formalism, the color-octet model provides a means to
bring theory into better agreement with data~\cite{Leibovich, ONIA1}.
The fundamental idea of this model is that while a $(c
\overline{c} )$ meson has to be in a color-singlet state, the
initially produced quark-antiquark pair does not.  One can produce,
for example, a $(c \overline{c} )$ pair in a color-octet $^3P$ state
which can then produce a color singlet $^3S_1$ $J/\psi$ meson by
single-gluon emission.  This is done at the cost of adding a small
number of parameters to the theory that currently must be determined
by experiment.  While the color-octet model can accommodate a large
cross section, strictly speaking it does not predict it. 
There are other deficiencies of the NRQCD formalism; for example, NRQCD 
expects that the spin alignment to be predominantly in the transverse  
state 
for the prompt $J/\psi$ mesons with large transverse momenta, 
a prediction that is not in agreement with 
subsequent measurement~\cite{Jpsi_pol}.

Previous prompt, direct, and inclusive $J/\psi$ cross-section
measurements~\cite{RUNIPSI} 
from CDF required a minimum transverse momentum
of 5\,GeV/$c$  on the $J/\psi$ although greater than 90$\%$ of the cross section has been
expected to lie below this point.  In this paper we present the
first measurement of the inclusive central $J/\psi$ cross section over
a much larger range of transverse momenta   from zero to 20\,GeV/$c$. The 
$J/\psi$ mesons are reconstructed from the decay channel 
 $J/\psi \rightarrow \mu^+ \mu^-$. 
The measurement was made possible by improving the CDF di-muon
trigger capability to be sensitive to $J/\psi$ with 
zero transverse momenta.

A significant fraction of $J/\psi$ mesons produced at the Tevatron come
from the decays of $b$-hadrons. In this experiment, we use the 
large  sample of $H_b \rightarrow J/\psi X$ events to measure the inclusive 
$b$-hadron cross section. The previous Tevatron measurements
\cite{CDFBXsec1, CDFBXsec2, CDFBXsec3, CDFBXsec4, D0BXsec1, D0BXsec2,
D0BXsec3} of the $b$-hadron cross section in proton-antiproton
collisions at $\sqrt{s}=1800$\,GeV were substantially larger (by a factor
of two to three) than  that predicted by next-to-leading order (NLO) QCD
calculations~\cite{BTHEORY1a,BTHEORY1b,Krey}. This was particularly
puzzling since the UA1 measurements at $\sqrt{s}=630$\,GeV~\cite{UA1BXsec} 
did not show such a marked departure from the
NLO QCD calculations. Several theoretical explanations were suggested:
higher-order corrections are large, intrinsic $k_T$ effects
are large~\cite{BTHEORY5}, extreme values of the renormalization
scales are needed, or new methods of resummation and fragmentation are
required~\cite{BTHEORY2, BTHEORY3, BTHEORY4}. Theories of new and
exotic sources of $b$-hadrons have also been proposed~\cite{SBOTTOM}. 
Since the  earlier Tevatron results covered only  10-13\% 
of the inclusive $p_T$ spectrum, 
it was not evident whether the excess was due to an overall increase
in the $b$-hadron production rate or a shift in the
spectrum toward higher $p_T$.

An inclusive measurement of $b$-hadron production over all transverse
momenta can help resolve this problem. Bottom hadrons have long
lifetimes, on the order of picoseconds~\cite{PDG}, which correspond to
flight distances of several hundred microns  at CDF.  We 
use the measured distance between the $J/\psi$ decay point and the beamline to
separate prompt production of charmonium from $b$-hadron decays.  
The single $b$-hadron cross section is then extracted from 
the measurement of the cross section of $J/\psi$ mesons 
from long-lived $b$-hadrons where single differentiates the cross section 
from the $b$-hadron cross section 
referring to $b$ and $\bar{b}$ hadrons which is a factor of two 
bigger.  In this paper, we present the first measurement of the inclusive single $b$-hadron
cross section at $\sqrt{s}=1960$\,GeV measured over all transverse
momenta in the rapidity range $|y|<0.6$.

\section{Description of the Experiment}
\subsection{The Tevatron}

The Fermilab Tevatron is a 1~km radius superconducting synchrotron.
Thirty-six bunches of 980\,GeV protons and antiprotons counter-circulate 
in a single ring and collide at
two interaction points (where the CDF and D0 detectors are located)
every 396~ns.  
The transverse profile of the interaction region can be 
approximately
described by a circular Gaussian distribution with a typical RMS width
of 30~$\mu \rm{m}$.   The longitudinal profile is also approximately 
Gaussian with a typical RMS of 30\,cm.  


For the data used in this analysis, instantaneous luminosities were in
the range $0.5$ to $2.0 \times 10^{31}$~$\rm{cm^{-2}  s^{-1}}$.  At
these luminosities, typically there was only a single collision 
in  a  triggered event.

\subsection{The CDF Detector}

In the CDF detector~\cite{CDFNIM1, CDFII}, a silicon 
vertex detector (SVX II)~\cite{SVX2Ref}, located immediately
 outside the beam pipe, provides
precise three-dimensional track reconstruction and is used to identify
displaced vertices associated with $b$ and $c$ hadron decays.  The
momentum of charged particles is measured precisely in the central
outer tracker (COT)~\cite{COT}, a multi-wire drift chamber that sits
inside a 1.4 T superconducting solenoidal magnet.  Outside the COT are
electromagnetic and hadronic calorimeters arranged in a
projective-tower geometry, covering the 
pseudo-rapidity region $|\eta|< 3.5$.   
Drift chambers and scintillator counters in the
region $|\eta| < 1.5$ provide muon identification outside the
calorimeters.
In the CDF coordinate system, $\theta$ and $\phi$ are the polar
and azimuthal angles, respectively, defined with respect to the proton beam
direction, $z$.  The pseudorapidity $\eta$ is defined as $- \ln
\tan(\theta/2)$.   The transverse momentum of a 
particle is $p_T = p \sin (\theta)$.

The portion of the silicon detector systems used in this analysis is
the SVX II detector. The SVX II consists of double-sided micro-strip
sensors arranged in five concentric cylindrical shells with radii
between 2.5 and 10.6\,cm. The detector is divided into 3 contiguous
five-layer sections along the beam direction for a total $z$ coverage 
of 90~\,cm. Each barrel is divided into
twelve azimuthal wedges of $30^{\circ}$ each. Each of the five
layers in a wedge is further divided into two electrically independent
modules called ladders. 
There are a total of 360 ladders in the SVX II detector.
The fraction of functioning ladders was increasing from 
78$\%$ to 94$\%$ during the period between  February 2002 and July 2002 
in which the data used in this paper were taken while the SVX detector 
was being commissioned.



The COT is the main tracking chamber in CDF. It is a cylindrical drift
chamber segmented into eight concentric superlayers filled with a
mixture of 50\% Argon and 50\% Ethane.  The active volume covers
$|z|<155$\,cm and 40 to 140\,cm in radius. Each superlayer is sectioned
in $\phi$ into separate cells. A cell is defined as one sense plane
with two adjacent grounded field sheets.  The sense plane is
composed of 40 $\mu$m gold-plated tungsten wires, twelve of which are
sense wires.  
In the middle of the sense planes, a mechanical spacer made of 
polyester/fiber glass rod is epoxied to each wire to limit
the stepping of wires out of the plane due to  electrostatic forces.
The main body of the field sheets
is 10 $\mu$m gold-coated mylar.
The field sheets approximate true grounded wire planes much better than
the arrays of wires which have often been used in wire chambers including the
predecessor to the COT.  Use of the field sheet also results in a smaller
amount of material within the tracking volume, and allows the COT to
operate at a much higher drift field than is possible with an array of
wires. The eight superlayers of the COT alternate between stereo and
axial, beginning with superlayer 1, which is a stereo layer. In an
axial layer, the wires and field sheets are parallel to the $z$ axis,
and thus provide only $r$-$\phi$ information. In stereo layers, the
wires and field sheets are arranged with a stereo angle of $\pm 2
^{\circ}$ and provide $z$ information in addition to $r$-$\phi$.

The CDF central muon detector (CMU)~\cite{CMU} is located around the
outside of the central hadron calorimeter at a radius of 347\,cm from
the beam axis.  The calorimeter is formed from 48 wedges, 24 on the east
(positive $z$), and 24 on the west (negative $z$), each wedge covering
15$^\circ$ in $\phi$. The calorimeter thickness is about 5.5 interaction
lengths for hadron attenuation.  The  muon drift cells with seven wires 
parallel to the beamline are  226\,cm long and
cover 12.6$^\circ$ in $\phi$. 
There is a 2.4$^\circ$ gap between drift cell arrays, giving a $\phi$
coverage of 84\%. The pseudorapidity coverage relative to the center of
the beam-beam interaction volume is $0.03<|\eta|<0.63$. Each wedge is
further segmented azimuthally into three 4.2$^\circ$ modules.  Each
module consists of four layers of four rectangular drift cells. The
sense wires in alternating layers are offset by 2 mm for ambiguity
resolution. The smallest unit in the CMU, called a stack, 
 covers about $1.2^\circ$ and includes 
four drift cells, one from each layer. Adjacent
pairs of stacks are combined together to form a two-stack unit called a
tower. A track segment detected in these chambers is called a CMU stub.

A second set of muon drift chambers is located behind an additional
60\,cm of steel (3.3 interaction lengths).  The chambers are 640\,cm
long and are arranged axially to form a box around the central
detector.  This system is called the CMP, and muons which register a
stub in both the CMU and the CMP 
are called CMUP muons.

Luminosity is measured using low-mass
gaseous Cherenkov luminosity counters (CLC)~\cite{LUMI1, LUMI2}. There are
two CLC modules in the CDF detector installed at small angles in the
proton and antiproton directions. Each module consists of 48 long,
thin conical counters filled with isobutane gas and arranged in
three concentric layers around the beam pipe. 

\subsection{Muon Reconstruction}

The starting point for the selection of $J/\psi \rightarrow \mu^+
\mu^-$ candidates is the reconstruction of two oppositely charged muons.
Muons are reconstructed from tracks measured in the tracking chambers
matched to the stub positions in the muon detectors.

\subsubsection{Charged Particle Tracking \label{sec_track}}
Track reconstruction begins in the COT.  The first step in the pattern
recognition is the formation of  line segments from hits in each
superlayer. Line segments from the axial layers that are tangent to a
common circle are linked together to form a track candidate and the
hit positions are fit to a circle. Line segments in stereo layers are
then linked to the 2-dimensional track and a helix fit is performed. The
transverse momentum resolution of the COT is measured using cosmic ray
events to be
\begin{equation}
\frac{\sigma({p_T})}{p_T^2} = 0.0017 \ [{\rm GeV/}c]^{-1}.
\end{equation}
The next step is to extrapolate each COT track into the SVX~II and add
hits that are consistent with the track. A window around the
track is established based on the errors on the COT track
parameters. If a hit in the outer SVX~II layer lies within the window,
it is added to the track. A new track fit is then performed, resulting
in a new error matrix and a new window. This window is then used to
add hits from the next SVX~II layer, and the procedure is repeated
over all layers. If no hit is found within the search window, the
algorithm proceeds to the next layer.  There may be multiple track
candidates with different combinations of SVX~II hits associated with
one COT track.  In this case, the track with the largest number of SVX
II layers with hits is chosen. A COT-SVX~II track is formed only if at
least three $r$-$\phi$ hits in the SVX~II are associated with the original COT
track.  An averaged impact parameter resolution  of 34~$\mu{\rm m}$ is 
achieved using hit information measured in  SVX~II for 
muon tracks  with $p_T$ around 1.5\,GeV/$c$.

\subsubsection{Muon Identification}
In the first stage of muon identification, the measured drift times for
hits in the muon chamber drift cells are converted to drift distances.  Hits 
in alternate layers that are within 7.5\,cm of each other are used to form
linear track segments. This distance corresponds to a maximum angle
relative to the radial direction in the chamber of $65^\circ$.  The
remaining pair of layers is then searched for hits within 0.5\,cm of
the line segment. The procedure is iterated and the optimal set of
hits is found. The segment resulting from a least-square fit to these
hits is called a ``stub". Hits are required  in at least 3 of the 4 layers  
to form a stub.


Stubs reconstructed in the CMU are matched to tracks with a minimal 
$p_T$ of 1.3~$\rm{GeV/c}$.
The tracks are extrapolated to the CMU after 
using a simplified geometry model to track the muon candidate's 
motion in the non-uniform magnetic field of the calorimeter. 
The distance,  $\Delta r\phi$, in the $r$-$\phi$ plane between the track
projected to the muon chambers and the muon stub is
required to be less than 30\,cm. The track is required to point to the
same side (east or west) of the detector that the stub is in unless
the muon candidate track  is within 20cm of the center of the detector.

\subsection{Triggers}
CDF uses a three-level trigger system~\cite{CDFII}.  At Level 1, data 
from every beam crossing is stored in a pipeline capable of buffering  
data from 42 beam-crossings. 
The Level 1 trigger either rejects the event or copies the data into one
of four Level 2 buffers.  During the data-taking period for this
analysis, the global Level 1 accept rate was approximately 10~kHz
corresponding to a rate reduction  factor of approximately 170.

At Level 2, a substantial fraction of the event data is available 
for analysis by the trigger processors which require approximately
25~${\rm \mu s}$ per event.   
During the period the data for this analysis were taken, the L2 accept 
rate was approximately 200~Hz, for a rejection factor of  
approximately 50.  

Events that pass the Level 1 and Level 2 selection criteria are then
sent to the Level 3 trigger~\cite{LEVEL3}, a cluster of computers running a 
speed-optimized reconstruction code. 
Events selected by Level 3 are
written to permanent mass storage.  During the period the data for
this analysis were taken, the global Level 3 accept rate was
approximately 40~Hz, for a rejection factor of approximately 5.

For the cross-section measurement, we require events with two muon
candidates identified by the Level 1 trigger.  
In Level 1, track
reconstruction is done by the eXtremely Fast Tracker
(XFT)~\cite{XFTcitation}.  The XFT examines COT hits from the
four axial superlayers and  provides $r$-$\phi$ tracking information.
The line segments  are identified in each superlayer and linked using 
predetermined patterns.  
 The XFT requires that each line segment contains hits found
on at least ten of a possible twelve anode wires in each axial superlayer. 
The XFT finds tracks with 
$p_T>1.5$\,GeV/$c$. It subdivides the COT into azimuthal sections of
$1.25^\circ$ each and places a track into a given section based on
its $\phi$ position at superlayer 6 ($r$ = 105.575\,cm).  If more than
one track candidate is found within a given section, the XFT return 
the track with the highest $p_T$.  The XFT passes the tracks it finds to the
eXTRaPolation unit (XTRP).  The XTRP extrapolates an XFT track's
trajectory to the CMU where a stub should be found if it
is a muon, taking into account the path of the track in the
magnetic field and the multiple scattering of
muon in the calorimeter.  
The XTRP then passes the search window to the
muon trigger crate, which looks for CMU stubs within
the search window.  A Level 1 CMU stub requires that there be hits
on both even layers or both odd layers of one $1.05^\circ$ stack of
the CMU with a drift time difference $\Delta t$  less than 396 ns.  
The twelve stacks in each $15^\circ$ wedge of the CMU are mapped in
pairs to six trigger towers to match the granularity of the XTRP
extrapolation.  If a muon stub is found within the search
window, it is considered a  Level 1 muon.  In order to fire the
di-muon trigger, two muon candidates must be found, separated by at least
two CMU trigger towers.  There is no requirement that the muons have
opposite charge at Level 1.  
During the data taking period in which the di-muon sample used for this 
analysis was obtained, there is no additional selection imposed on muons 
at Level 2 and event is passed to Level 3 directly from Level 1. 



At Level 3, the muons are required to have opposite charge,
and to have an invariant mass between 2.7 and 4.0\,GeV/$c^2$. 
In addition, both muon tracks are required to be within 
5~\,cm in $z_0$,  where $z_0$ is 
the $z$ coordinate of the muon track at its distance of the closest approach 
in the $r$-$\phi$ plane to the beam axis. 
For a portion of the data sample considered in
this analysis, there is a requirement that the opening angle in
$r$-$\phi$ between the di-muons be less than $130^{\circ}$.


\subsection{Luminosity}

The CLC counters monitor the average number of inelastic $p\bar{p}$
interactions in each bunch crossing.  The inelastic $p\bar{p}$ cross
section has been measured to be $\sigma_{\rm in}\sim 60 $ mb by several
experiments at $\sqrt{s}=1800$\,GeV~\cite{PPXSEC1, PPXSEC2,
PPXSEC3}. The inelastic $p\bar{p}$ cross section at $\sqrt{s}=1960$\,GeV is 
scaled from previous measurements using the calculations in~\cite{BLOCK}.  
The rate of inelastic $p\bar{p}$
interactions is given by
\begin{equation}
\mu \cdot f_{\rm BC} = \sigma_{in} \cdot {\cal L},
\label{eqn_lumi}
\end{equation}
where ${\cal L}$ is the instantaneous luminosity, $\mu$ is the average number
of inelastic  $p\bar{p}$ interactions per bunch crossing, and $f_{\rm BC}$ is the
rate of bunch crossings. In this paper, we use data from the beginning
of the CDF Run~II operation where the average instantaneous luminosities
were relatively low. 

The number of $p\bar{p}$ interactions in a bunch crossing follows
Poisson statistics where the probability of empty
crossings is given by $\mathcal{P}_0(\mu) = e^{-\mu}$.
An empty crossing is observed when there are fewer than two counters with
signals above threshold in either module of the CLC.  The measured
fraction of empty bunch crossings is corrected for the CLC acceptance
and the value of $\mu$ is calculated. The measured value of $\mu$ is
combined with the inelastic $p\bar{p}$ cross section to determine the 
instantaneous luminosity using Equation~\ref{eqn_lumi}.  
Because this method depends only weakly on
the CLC thresholds, it functions particularly well at low 
luminosities where the probability of empty bunch crossings is large.  
The systematic error on the luminosity measurement is estimated to be $6\%$.

In CDF Run~II, only runs with greater than 10~nb$^{-1}$
integrated luminosity are considered for analysis. 
Runs with good operating conditions in the detector 
are tagged by the online shift crews.  
Data from those runs are examined to exclude ones with COT,
muon or trigger hardware problems. 
For the measurement presented in this paper, the data 
collected from February to July 2002 was used. 
This sample corresponds to a total
integrated luminosity of $39.7 \pm 2.3$ pb$^{-1}$.

For $J/\psi$ candidates with transverse momenta in the range 0 to 2
\,GeV/$c$, we use $14.8 \pm 0.9$ pb$^{-1}$ of our data sample, which
corresponds to that fraction of the data collected when no cut on the
di-muon opening angle in the Level 3 trigger was used.  

\section{Data Selection and Reconstruction}
\subsection{Data selection}
The events selected by the $J/\psi$ trigger are reconstructed offline, taking 
advantage of the most refined constants and algorithms. 
We reconstruct $J/\psi \rightarrow \mu^+ \mu^-$ decays by selecting
events with two oppositely charged muon candidates reconstructed in the
COT and CMU detectors.  The $J/\psi \rightarrow \mu^+ \mu^-$ sample
used for this analysis was collected using the CMU di-muon
triggers. Events are required to have satisfied the Level 1 and Level
3 di-muon trigger criteria.



\begin{figure}[!h]
\centerline{\psfig{figure=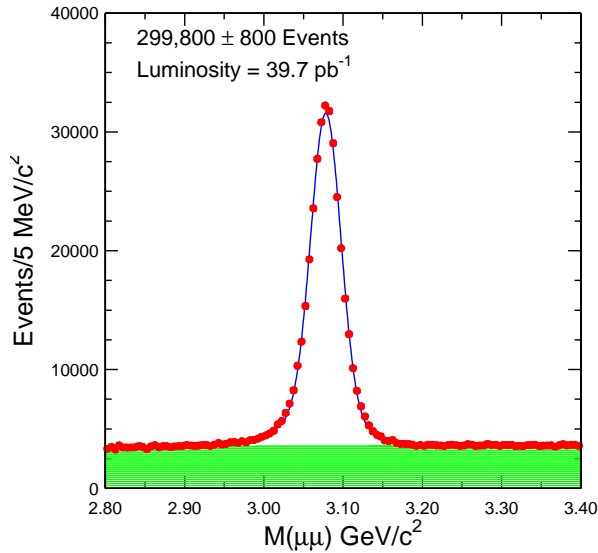,width=0.5\textwidth}}
\caption{Mass distribution of reconstructed di-muon
$J/\psi$ candidates.  The points are data. 
The solid line is the fit to the signal approximated 
as a double Gaussian and a linear fit for the background. The hatched region 
is the fitted background. 
The fit gives a signal of  $299800\pm 800$ $J/\psi$ events with an averaged 
mass of  $3.09391\pm0.00008$\,GeV/$c^2$ obtained  
and an average width of $0.020\pm0.001$\,GeV/$c^2$ mainly due to 
detector resolution. The uncertainties here  are statistical only.
}
\label{jpsi-rawmass.eps}
\end{figure}

In addition to the default muon selection criteria outlined
earlier, we require a $p_T$ independent track-stub matching criterion
$\chi^2({\Delta r\phi})<9$.  
A track-stub matching quality criterion $\chi^2({\Delta r\phi})$ with a one
degree of freedom  is
calculated from $\Delta{r\phi}$ and the expected multiple scattering
for a track of given $p_T$ obtained from a GEANT simulation~\cite{GEANT} of 
the CDF Run II  detector material. 
We require both muons to have transverse
momenta $p_T>1.5~{\rm\,GeV}/c$ as measured offline.  
The trigger requirements are verified for the offline-reconstructed 
candidates.
In addition, each CMU stub matched to a triggered stub must
lie within the XTRP search window set by the Level 1
triggered track.
Furthermore, track momentum is corrected for 
energy loss due to specific ionization and multiple scattering
according to our accounting of the detector materials.  
We calculate the $J/\psi$ candidate invariant mass from the
four-momenta of the two muons.
For a portion of the data sample under consideration, a temporary 
hardware problem with the  
di-muon logic caused the trigger to exclude $J/\psi$ events where
both muon stubs fell in the $\phi$ range of
$240-270^\circ$.  Therefore, we exclude $J/\psi$ events where both
muons fall in that $\phi$ region and account  for this in the detector
acceptance. 
We also reject $J/\psi$ candidates if one of 
the tracks passes within 1.5~\,cm of the center of any  
COT wire planes,  where the trigger efficiency is difficult to model 
because of the distortion of the electric field due to the mechanical spacers.
This exclusion is accounted for in the acceptance
calculation. The muon reconstruction efficiency is measured in each of
the 48 CMU detector wedges. We find that the hit efficiency in the CMU
wedge on the west side of the detector covering the region $240^\circ<
\phi< 255^\circ$ is lower due to a known hardware problem and  
exclude $J/\psi$ events
where either muon stub is reconstructed in this wedge.  
As shown in Fig.~\ref{jpsi-rawmass.eps},
there are  $299800\pm 800$ $J/\psi$ events that 
passed these selection conditions.


\begin{figure}[!h]
\centerline{
\psfig{figure=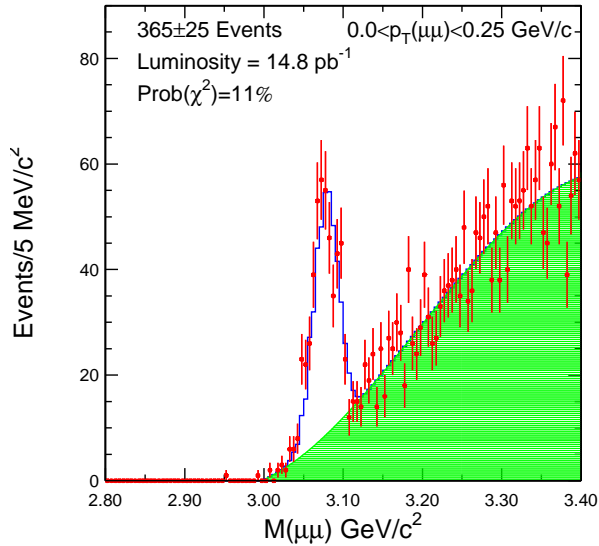,width=0.5\textwidth}}
\caption{Invariant mass distribution of reconstructed $J/\psi
\rightarrow \mu \mu$ events in the range $p_T(\mu\mu)<0.25$\,GeV/$c$. The points with error bars are data. The solid line is
the fit to the signal shape from the simulation and a third order
 polynomial for the background. The shaded histogram is the
fitted background shape.  The number of signal events and the fit 
probability of the binned $\chi^2$ fitting are also provided.}
\label{fig_mass1a}
\end{figure}

\begin{figure}[!h]
\centerline{
\psfig{figure=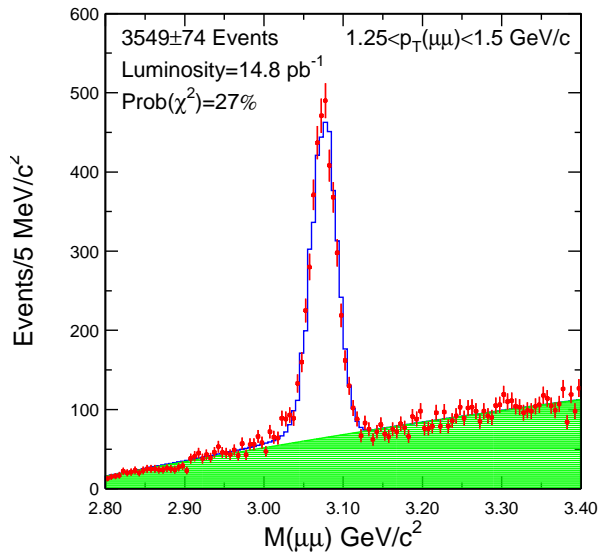,width=0.5\textwidth}}
\caption{Invariant mass distribution of reconstructed $J/\psi
\rightarrow \mu \mu$ events in the range $1.25<p_T(\mu\mu)<1.5$\,GeV/$c$. The points with error bars are data. The solid line is
the fit to the signal shape from the simulation and a second order
 polynomial for the background. The shaded histogram is the
fitted background shape.}
\label{fig_mass1d}
\end{figure}

\begin{figure}[!h]
\centerline{
\psfig{figure=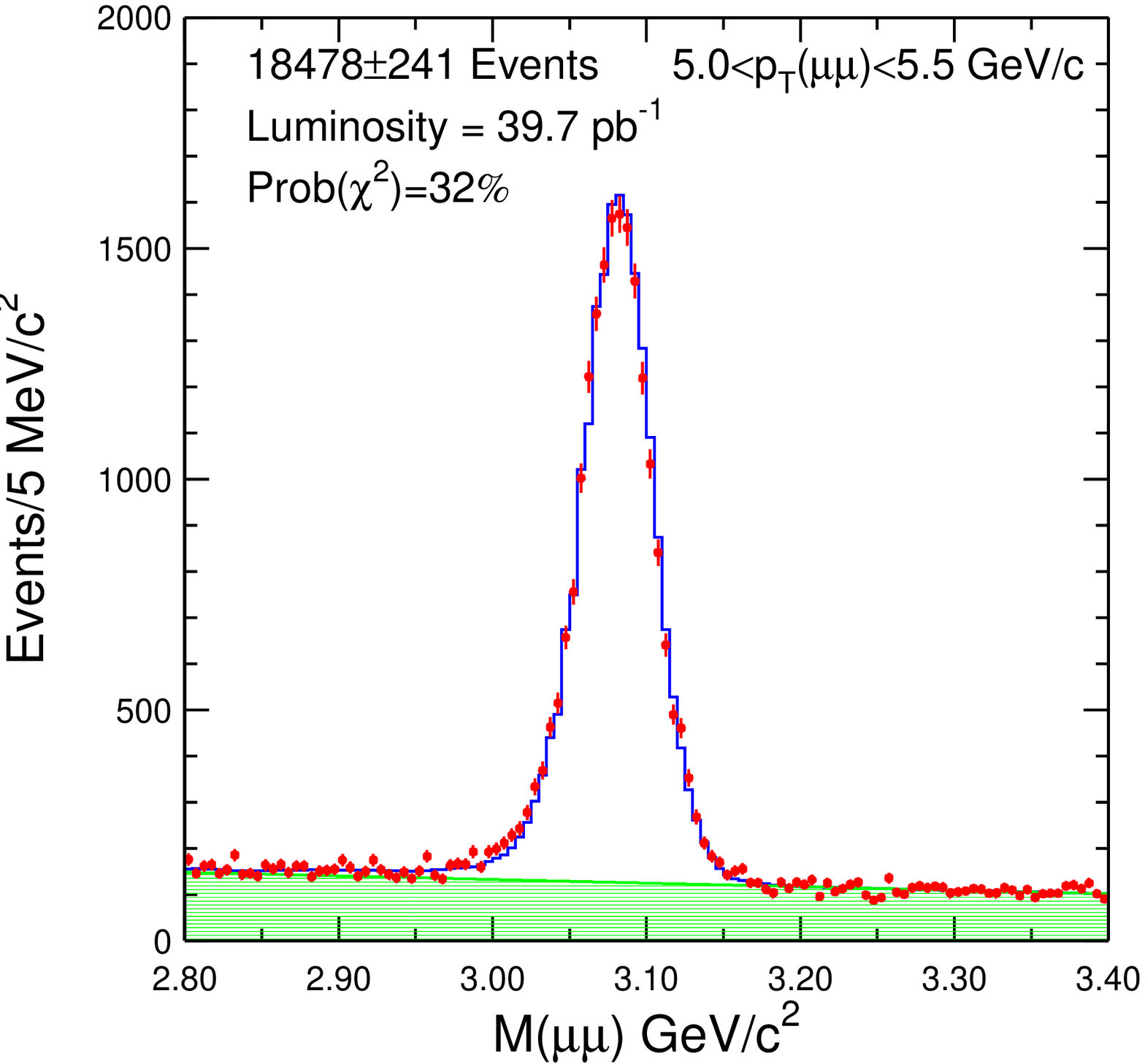,width=0.5\textwidth}
}
\caption{Invariant mass distribution of reconstructed $J/\psi
\rightarrow \mu \mu$ events in the range $5.0<p_T(\mu\mu)<5.5$\,GeV/$c$. The points with error bars are data. The solid line is
the fit to the signal shape from the simulation and a linear 
background. The shaded histogram is the
fitted background shape.}
\label{fig_mass1b}
\end{figure}

\begin{figure}[!h]
\centerline{
\psfig{figure=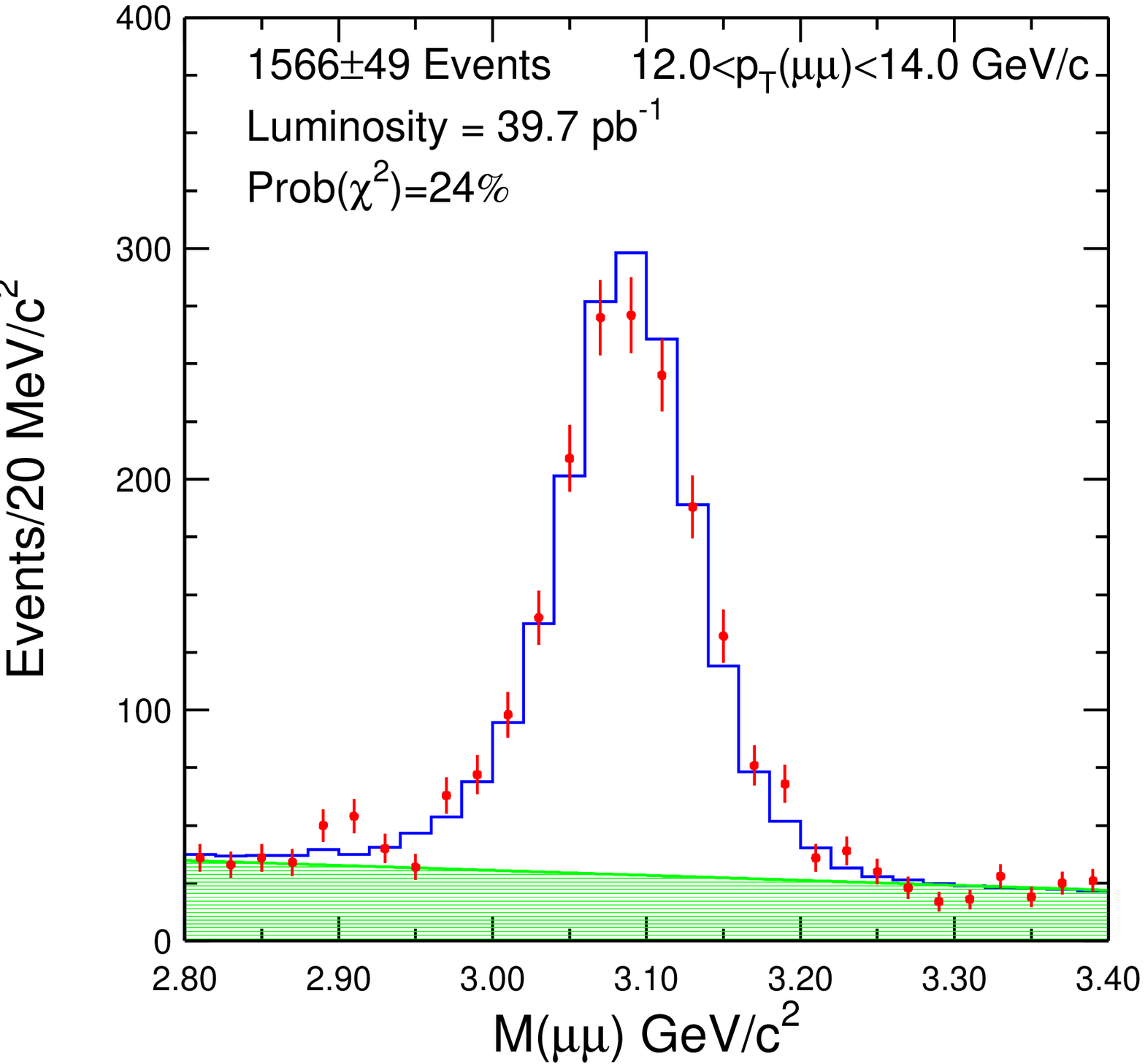,width=0.5\textwidth}
}
\caption{Invariant mass distribution of reconstructed $J/\psi
\rightarrow \mu \mu$ events in the range $12.0<p_T(\mu\mu)<14.0$\,GeV/$c$. The points with error bars are data. The solid line is
the fit to the signal shape from the simulation and a linear background.
The shaded histogram is the fitted background shape.}
\label{fig_mass1c}
\end{figure}

To determine the yield in each $J/\psi$ $p_T$ bin, the di-muon
invariant mass distributions  are fitted 
using invariant mass line shapes  including the radiative tail from internal
bremsstrahlung obtained from a tuned hit-level COT simulation. 
The simulated $J/\psi$ events
are decayed using the $J/\psi$ radiative decay model in the QQ decay
package~\cite{QQ}.  The COT hit multiplicity per track is tuned to
match the data as closely as possible. The COT hit resolution is then
tuned to find the best $\chi^2$ in a binned fit to the data
using the Monte Carlo invariant mass line shape for the signal and a
polynomial shape for the background. Finally, energy loss and multiple
scattering in material encountered before the COT are modeled. 
The energy loss in the silicon
material is scaled until the peaks of the di-muon invariant mass
distribution in different $p_T$ ranges in data and from the simulation
match. 
The order of the
background polynomial used varies with the background shape in each
$J/\psi$ $p_T$ range. A third-order polynomial is used for the
momentum range 0-0.25\,GeV/$c$,  a second-order polynomial is used for
the range 0.25-2.25\,GeV/$c$,  and a
first-order polynomial (linear background) for 
transverse momenta greater than $2.25$\,GeV/$c$.  
The fits to the invariant mass distributions in 
four $J/\psi$ $p_T$ ranges are shown in 
Figs.~\ref{fig_mass1a} - ~\ref{fig_mass1c}.
The $J/\psi$  yields and the statistical uncertainties obtained from the fits 
in  each $p_T$ range  are listed in  the first column of 
Table~\ref{tab_xsecsum}. 
The mass fitting qualities over the whole $p_T$ bins are 
good as indicated from the fit probability shown in these Figures. 
We also examined the differences between counting the event numbers 
in the $J/\psi$ signal region ( $3.02 \rightarrow 3.15$\,GeV/$c^2$) 
to that predicted from the fitting functions of signal and background. 
The differences ranging from  +9$\%$ in the 
lowest $p_T$ bin to -1.3$\%$ in the high $p_T$ bin are used 
very conservatively  as the systematic uncertainties 
from the mass fitting.

\begingroup
\squeezetable
\begin{table*}
\caption{
Summary of the inclusive $J/\psi$ cross-section analysis components. The values of the
yield and statistical uncertainty from the fits are listed in the 2nd
column. The acceptance values and the combined systematic and
statistical uncertainties on the acceptance are listed in the 3rd
column. In the 4th and 5th columns the trigger and track-stub
matching efficiencies obtained from the mean of the distribution in
each bin and the corresponding systematic
uncertainties are listed. The sixth column lists the 
integrated luminosity used for each measurement. 
}
\begin{center}
\begin{tabular}{cccccc}  
\hline \hline
$P_t$ range  & Yield           & Acceptance    & Level 1 Trigger & Track-stub  matching & Luminosity\\
\,GeV/$c$       &  ($N^i$)               & ($\mathcal{A}^i$) &  Efficiency
($\epsilon_{L1}^i$) &
Efficiency($\epsilon_{\chi^2}^i$) & ($\mathcal{L}^i$) nb$^{-1}$\\ \hline 
$0.0-0.25$   &$365\pm     25$  & $ 0.0153  \pm 0.0007  $ & $ 0.857 \pm 0.013 $ &$  0.9963 \pm 0.0009$ &  $14830 \pm 870$ \\
$0.25-0.5$   &$605\pm     30$  & $ 0.0069 \pm 0.0004  $ & $ 0.860 \pm 0.013 $ &$  0.9963 \pm 0.0009$ & "  \\
$0.5-0.75$   &$962\pm     38$  & $ 0.0070 \pm 0.0004  $ & $ 0.865 \pm 0.013 $ &$  0.9962 \pm 0.0009$ & "  \\
$0.75-1.0$   &$1592\pm    49$  & $ 0.0087 \pm 0.0005  $ & $ 0.871 \pm 0.014 $ &$  0.9961 \pm 0.0009$ & "  \\
$1.0-1.25$   &$2500\pm    62$  & $ 0.0116 \pm 0.0006  $ & $ 0.877 \pm 0.014 $ &$  0.9960 \pm 0.0009$ & "  \\
$1.25-1.5$   &$3549\pm    74$  & $ 0.0151 \pm 0.0008  $ & $ 0.885 \pm 0.014 $ &$  0.9957 \pm 0.0009$ & "  \\
$1.5-1.75$   &$4517\pm    84$  & $ 0.0190 \pm 0.0009  $ & $ 0.892 \pm 0.014 $ &$  0.9955 \pm 0.0009$ & "  \\
$1.75-2.0$   &$5442\pm    93$  & $ 0.0232 \pm 0.0011  $ & $ 0.899 \pm 0.015 $ &$  0.9953 \pm 0.0009$ & "  \\
$2.0-2.25$   &$16059\pm  167$  & $ 0.0271 \pm 0.0013  $ & $ 0.905 \pm 0.015 $ &$  0.9960 \pm 0.0009$ & $39700 \pm 2300 $   \\
$2.25-2.5$   &$18534\pm  252$  & $ 0.0317 \pm 0.0015  $ & $ 0.911 \pm 0.015 $ &$  0.9946 \pm 0.0009$ & "  \\
$2.5-2.75$   &$18437\pm  253$  & $ 0.0367 \pm 0.0017  $ & $ 0.916 \pm 0.015 $ &$  0.9943 \pm 0.0009$ & "  \\
$2.75-3.0$   &$18858\pm  259$  & $ 0.0415 \pm 0.0019  $ & $ 0.920 \pm 0.015 $ &$  0.9939 \pm 0.0009$ & "  \\
$3.0-3.25$   &$18101\pm  253$  & $ 0.0467 \pm 0.0021  $ & $ 0.924 \pm 0.015 $ &$  0.9935 \pm 0.0009$ & "  \\
$3.25-3.5$   &$17597\pm  250$  & $ 0.0532 \pm 0.0024  $ & $ 0.927 \pm 0.015 $ &$  0.9931 \pm 0.0009$ & "  \\
$3.5-3.75$   &$16400\pm  241$  & $ 0.0576 \pm 0.0025  $ & $ 0.930 \pm 0.015 $ &$  0.9927 \pm 0.0009$ & "  \\
$3.75-4.0$   &$14863\pm  226$  & $ 0.0628 \pm 0.0029  $ & $ 0.932 \pm 0.015 $ &$  0.9923 \pm 0.0009$ & "  \\
$4.0-4.25$   &$14056\pm  218$  & $ 0.0694 \pm 0.0031  $ & $ 0.934 \pm 0.015 $ &$  0.9918 \pm 0.0010$ & "  \\
$4.25-4.5$   &$12719\pm  212$  & $ 0.0768 \pm 0.0034  $ & $ 0.936 \pm 0.015 $ &$  0.9913 \pm 0.0010$ & "  \\
$4.5-4.75$   &$12136\pm  201$  & $ 0.0840 \pm 0.0037  $ & $ 0.937 \pm 0.014 $ &$  0.9909 \pm 0.0010$ & "  \\
$4.75-5.0$   &$10772\pm  188$  & $ 0.0904 \pm 0.0039  $ & $ 0.939 \pm 0.014 $ &$  0.9904 \pm 0.0010$ & "  \\
$5.0-5.5$    &$18478\pm  241$  & $ 0.1006 \pm 0.0042  $ & $ 0.940 \pm 0.014 $ &$  0.9897 \pm 0.0010$ & "  \\
$5.5-6.0$    &$14616\pm  210$  & $ 0.1130 \pm 0.0046  $ & $ 0.942 \pm 0.014 $ &$  0.9887 \pm 0.0011$ & "  \\
$6.0-6.5$    &$11388\pm  180$  & $ 0.1257 \pm 0.0051  $ & $ 0.946 \pm 0.014 $ &$  0.9876 \pm 0.0011$ & "  \\
$6.5-7.0$    &$8687\pm   154$  & $ 0.1397 \pm 0.0055  $ & $ 0.945 \pm 0.014 $ &$  0.9865 \pm 0.0012$ & "  \\
$7.0-8.0$    &$12409\pm  139$  & $ 0.1561 \pm 0.0068  $ & $ 0.946 \pm 0.014 $ &$  0.9850 \pm 0.0012$ & "  \\
$8.0-9.0$    &$6939\pm   107$  & $ 0.1723 \pm 0.0075  $ & $ 0.947 \pm 0.014 $ &$  0.9827 \pm 0.0013$ & "  \\
$9.0-10.0$   &$3973\pm    78$  & $ 0.1807 \pm 0.0079  $ & $ 0.948 \pm 0.014 $ &$  0.9804 \pm 0.0014$ & "  \\
$10.0-12.0$  &$3806\pm    74$  & $ 0.1938 \pm 0.0074  $ & $ 0.949 \pm 0.014 $ &$  0.9772 \pm 0.0016$ & "  \\
$12.0-14.0$  &$1566\pm    49$  & $ 0.2163 \pm 0.0081  $ & $ 0.960 \pm 0.014 $ &$  0.9726 \pm 0.0017$ & "  \\
$14.0-17.0$  &$935\pm     40$  & $ 0.238  \pm 0.011   $ & $ 0.951 \pm 0.014 $ &$  0.9671 \pm 0.0018$ & "  \\
$17.0-20.0$    &$350\pm     25$  & $ 0.247  \pm 0.012   $ & $ 0.951 \pm 0.014 $ &$  0.9600 \pm 0.0020$ & "  \\
\hline \hline
\end{tabular}
\label{tab_xsecsum}
\end{center}
\end{table*}
\endgroup



\section{Acceptance and Efficiency}
\subsection{Monte Carlo Description  \label{sec_mc}}

We use the $\small {\rm GEANT}$~\cite{GEANT} Monte Carlo simulation software to
estimate the geometric and kinematic acceptances. The variation of
detector conditions in the simulation is set to match the data. 
$J/\psi$ events are generated starting with a kinematic
distribution that is flat in rapidity and with a $p_T$ distribution
selected to best match the reconstructed data. The events are fully
simulated. After the differential cross section  is measured, we iterate 
 and recalculate the acceptance and the central value of the
cross section using the measured $p_T$ distribution. 
The $\small {\rm GEANT}$ simulation is validated by comparing the 
resulting distributions of
various kinematic quantities such as $\eta$, $p_T$, 
the track-stub matching distance, and 
the $z$ vertex distribution in reconstructed
data and reconstructed Monte Carlo events. Differences in the data and
Monte Carlo distributions are used to estimate the systematic
uncertainties on the modeling of the CDF detector geometry in the
simulation.

\subsection{Acceptance}


We correct the observed number of $J/\psi$ events for the detector acceptance 
and  efficiency.
The CMU muon detector covers the pseudo-rapidity range of
$\mid \eta \mid<0.6$.
In this
region the coverage of the COT is complete and
the CDF detector acceptance is driven by the muon detector geometry
and kinematic reach. The calorimeter acts as an absorber for the CMU
detector which is therefore sensitive only to muons with $p_T > 1.35$
\,GeV/$c$. The arrangement of the four sense wires within the CMU chambers
allows a lower bound on the transverse momentum of the muon to be
calculated from the difference in drift times in sense wires on
alternating layers. The $\Delta t \leq 396$ ns timing window is
selected to be fully efficient for muons with $p_T>1.5$\,GeV/$c$.



\begin{figure}[!h]
\psfig{figure=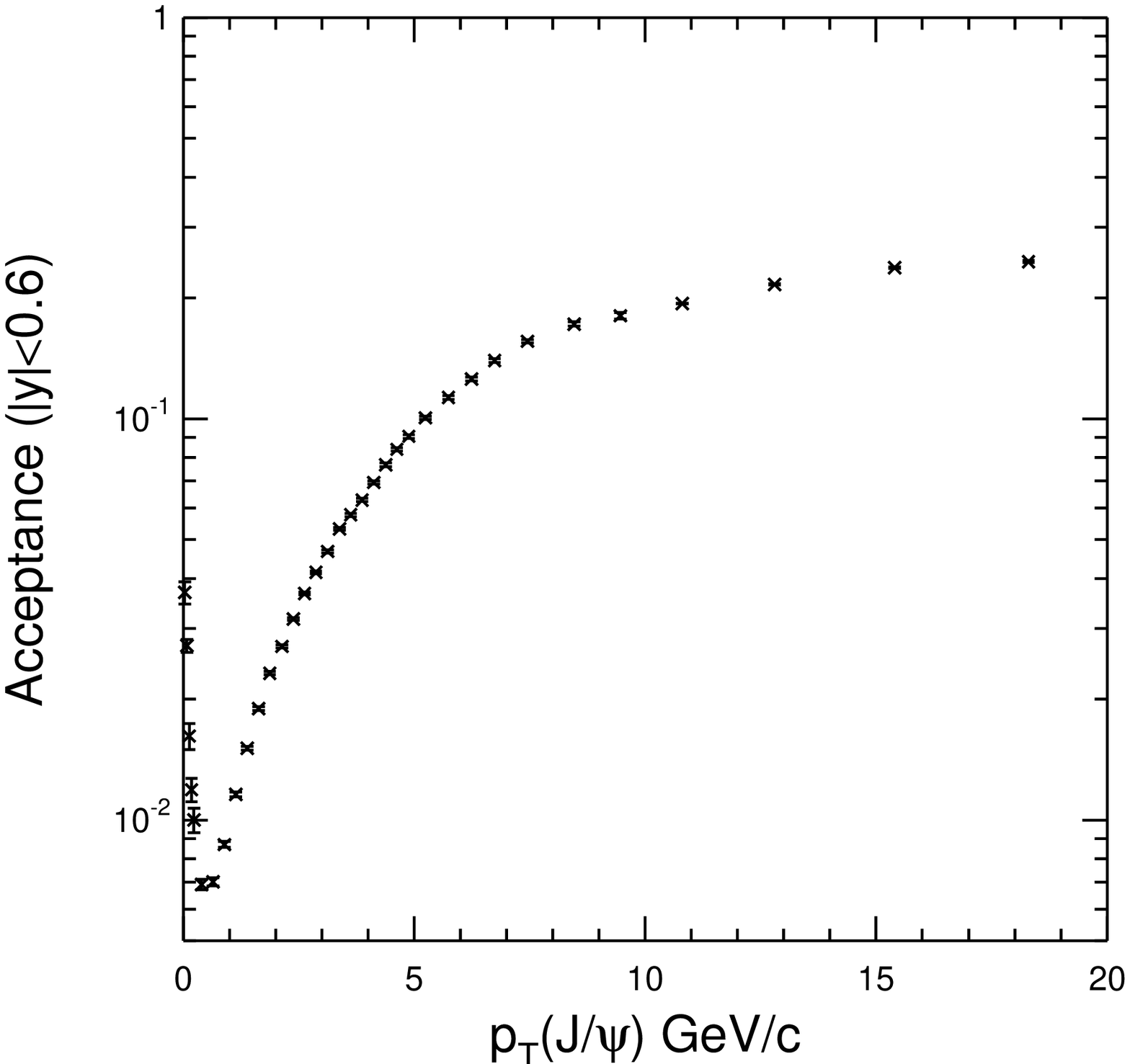,width=0.5\textwidth}
\psfig{figure=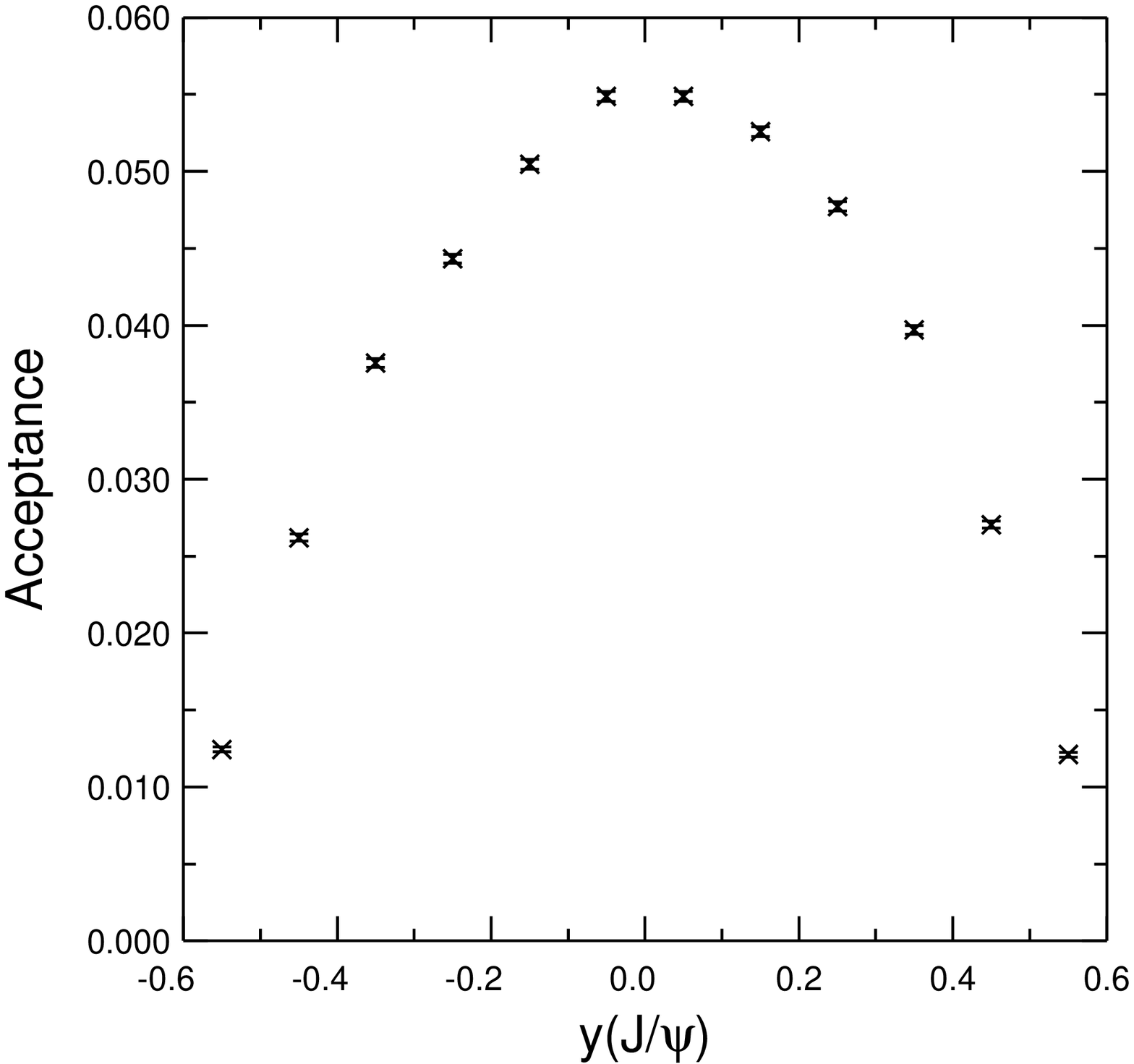,width=0.5\textwidth}
\caption{ Acceptance of $J/\psi \rightarrow \mu \mu$  events 
determined from a GEANT simulation of the CDF detector. The
acceptance is shown as a function of $p_T(J/\psi)$ and
$y(J/\psi)$. The acceptance as a function of $p_T(J/\psi)$ is measured
integrated over $|y|<0.6$ and the acceptance as a function of $y$ is
shown integrated over all $p_T$. 
}
\label{fig_491acc2}
\end{figure}

The acceptance is modeled as a function of both the 
reconstructed $p_T(J/\psi)$ and rapidity $y(J/\psi)$ and 
is defined as the ratio between the number of generated events $N^{gen}$
 and recontructed events $N^{rec}$,
\begin{equation}
{\mathcal{A}}(p_T,y) =
\frac{N^{rec} (p_T(J/\psi),|y(J/\psi)|<0.6) }
 {N^{gen}(p_T'(J/\psi),|y'(J/\psi)|<0.6)}, 
\end{equation}
where $p_T'(J/\psi)$ and $y'$ are the generated true values of the
$J/\psi$ momentum and rapidity.  
The acceptance as a function of $p_T$ and $y$ is shown in 
Fig.~\ref{fig_491acc2}.

The acceptance increases rapidly from 0.7$\%$ at $p_T = 0.25$\,GeV/$c$ to
10$\%$ at 5\,GeV/$c$ and 25$\%$ at 20\,GeV/$c$. 
The acceptance in the range 0.0-0.25\,GeV/$c$ is rapidly 
varying as a function of $p_T(J/\psi)$ 
and increases with decreasing momenta from 0.7$\%$ at $p_T(J/\psi)
=0.25$\,GeV/$c$ to 4\% for $J/\psi$ mesons almost at rest 
($p_T < 50~$MeV/$c$).  The muon transverse momentum is required to be greater
than or equal to 1.5\,GeV/$c$, which is close to one-half of the
$J/\psi$ mass, therefore when the $J/\psi$ is at rest both muons are
likely to be above the $p_T$ threshold. As soon as the $J/\psi$
receives a small boost, the probability is greater that at least one
muon will be below the $p_T$ acceptance threshold and the acceptance
starts to decrease until the $J/\psi$ transverse momentum exceeds 0.25\,GeV/$c$.

There is a small but
non-zero acceptance at $|y| = 0.6$ due to detector resolution and the
size of the interaction region.  $J/\psi$ Monte Carlo events 
generated with a
flat rapidity distribution in the range $|y|_{\rm gen}<1.0$ and a $p_T$
distribution as described in Section~\ref{sec_mc}
are simulated. The relative acceptance of events
generated with $|y|_{\rm \rm gen}>0.6$ and reconstructed with
$|y|_{\rm \rm reco}<0.6$, ${\mathcal{A}}'$, is calculated thus:
\begin{equation}
 {\mathcal{A}}' = \frac{N^{\rm rec}(|y|_{\rm gen}>0.6,|y|_{\rm
rec}<0.6)} {N^{gen} (|y|_{\rm gen}<0.6)},
\label{eqn_ysmear}
\end{equation}
where $ N^{\rm rec}(|y|_{\rm gen}>0.6,|y|_{\rm rec}<0.6)$ is the
number of $J/\psi$ events in the Monte Carlo sample with reconstructed $|y|_{\rm \rm
reco} < 0.6$  and generated $|y|_{\rm gen} > 0.6$ and $N^{gen} (|y|_{\rm
gen}<0.6)$ is the total number of events generated with $|y|<0.6$. 
The value of ${\mathcal{A}}'$ is found to be 
very small: ${\mathcal{A}}' = 0.00071
\pm 0.00006(stat)$. A correction factor of $(1-{\mathcal{A}}') =
99.93\%$ is applied to the $J/\psi$ yield calculated in each
$p_T(J/\psi)$ bin.


A 2-dimensional  acceptance function was  used for an 
event-by-event correction during the cross section calculation process. 
In Table~\ref{tab_xsecsum}, the averaged acceptance values and 
the combined statistical and systematic uncertainties for  
each $p_T$ bin are given. 
Sources of  systematic uncertainties studied are 
$J/\psi$ spin alignment, $p_T$ spectrum, CMU simulation and 
detector material description in GEANT simulation.

Kinematic acceptance as a function of $p_T$ depends on the $J/\psi$
spin alignment. The normalized alignment distribution is given by
\begin{equation}
I(\theta) = \frac{3}{2(\alpha+3)} (1 +
\alpha \cos^2\theta), 
\end{equation}
where $\theta$ is the angle between the muon in the $J/\psi$ rest
frame and the direction of the $J/\psi$ in the lab 
frame~\cite{Jpsi_pol} and $\alpha$ quantifies the spin alignment. The
parameter $\alpha$ must lie in the range -1 to 1 and $\alpha=0$
indicates no preferred spin alignment.   
The previous CDF measurements of
the $J/\psi$ spin alignment parameter~\cite{Jpsi_pol} are consistent
with zero but could also be as large as $50\%$ in some $p_T$
regions. The weighted mean of $\alpha$ measured in different $p_T$
ranges in~\cite{Jpsi_pol} is used to determine the central value of
the parameter $\alpha$ to be used for the acceptance. The value of
$\alpha=0.13 \pm 0.15$ is used for the final acceptance values where
the uncertainty is chosen to accommodate the variation in the previous
CDF measurements and the extrapolation to $p_T=0$ where $\alpha$ is
expected to be zero. The uncertainty on acceptance 
due to spin alignment is largest
in the lower momentum bins and decreases with increasing transverse
momentum. We find the uncertainty is $\sim 5\%$ near $p_T=0$ and $2\%$
in the region $17<p_T<20$\,GeV/$c$.

To estimate the uncertainty from variations of the input 
transverse momentum spectrum,  the acceptance is recalculated using
a Monte Carlo sample generated using a  flat distribution. 
The flat distribution is an extreme alternative from 
the nominal spectrum  which is a  fast falling function of $p_T$.
The fractional change in acceptance is taken
as the uncertainty on the input transverse momentum
distribution.  The uncertainty is about 3\% 
in the lowest momentum bin, 
less than $1\%$ in the 0.25 to 3\,GeV/$c$ bins,
1-2$\%$ in the 3 to 4\,GeV/$c$ bins, 
and 2-4$\%$ in the 4 to 20 GeV/$c$ bins.

A systematic error of $1.0\%$ from uncertainties related to the  CMU 
chamber simulation is estimated by  comparing event
distributions in data and in Monte Carlo. 
The modeling  of the CMU coverage in $r$-$z$ plane,  
the wire efficiency differences between wedges in east and west 
and in different $\phi$ sections, and beam position in $z$ are found to be 
the major sources of the simulation uncertainties.

There is a gap in CMU coverage in the central region of the
detector in the $r$-$z$ plane. The gap in coverage is approximately $\pm
11$\,cm,  measured at a radius of $347$\,cm. 
The fraction of muons falling in the gap region but still accepted by
the CMU due to multiple scattering is compared between data and Monte
Carlo. The deviation between the ratios in data and Monte Carlo is
taken as the uncertainty in the modeling of the CMU fiducial volume
in the center of the CDF detector. The uncertainty is found to be  
$0.20\%$.


Several factors contribute to the difference in the numbers of $J/\psi$ mesons 
with decay vertex  in the opposite halves of the detector along $z$. 
These include the shift in the average primary vertex location 
towards positive $z$ (east), the exclusion of the low efficiency wedge on the
west side of the detector, and the uncertainty in the modeling of
the $z$ extent of the CMU detector, as well as the differences in the
east and west chambers.  We found a  difference of $0.80\%$ between data 
and Monte Carlo on the east-west asymmetry in the number of reconstructed 
$J/\psi$ events. 

The $\phi$ acceptance of the CMU detector obtained from the GEANT
simulation does not include the differences in gain and
efficiencies between wedges. The number of events reconstructed in
each wedge in data and Monte Carlo is examined and the total number of
events in Monte Carlo is normalized to match the data. The standard
deviation of the difference between the number of events reconstructed in
each wedge between data and Monte Carlo is taken as the uncertainty on the
CMU $\phi$ acceptance. We find an  uncertainty of
$0.55\%$ due to this source.

Muons from $J/\psi$ are required to have the $z_0$ position 
to be within  90~\,cm of the center of the detector, 
$|z_0| < 90$~\,cm. 
There is a small disagreement
between the data and the Monte Carlo in the 
$z_0(\mu)$ distributions due to inadequate modeling of the 
interaction region. This contribution to
the systematic error is estimated from the difference between  the ratios of
data and Monte Carlo tracks with $|z_0|<90$\,cm compared to all
muons. We find an uncertainty of $0.28\%$.

The material description of the CDF detector in GEANT 
determines the amount of energy loss from a muon track when it travels 
through the detector.   
Inside the tracking volume, the material description of the  new 
 silicon detector has the biggest impact  on muon tracks in the 
low momentum range which is of special interest to this analysis.
To estimate the systematic error on the acceptance from uncertainty of
the detector material description,   the SVX II
material used in the simulation was varied by  10~$\%$ to 20$\%$.  
The systematic uncertainty is taken as the difference between the
acceptance values measured with different material scale factors and
the nominal.  The uncertainty is largest in the low momentum bins
where it is around 5\%.

The systematic uncertainties on acceptance are summarized 
in Table~\ref{tab_syst}. The size of the uncertainties from $J/\psi$ spin 
alignment, $J/\psi$ $p_T$ spectrum and detector material 
description depends on the  muon $p_T$ range as  
expected while the uncertainty from muon detector simulation 
is same for all $p_T$ ranges of interests in the analysis. 

\subsection{Data Quality}

The yield, mean, and resolution of the $J/\psi$ invariant mass peak were 
monitored over the period of the data taking to evaluate the detector 
performance.  
The number of $J/\psi$ mesons reconstructed is
normalized by the integrated luminosity of each run.
We identify outlying runs
which may have additional hardware or trigger problems that have been
undetected by the standard offline validation procedures. Runs with
$J/\psi$ yields different by $4 \sigma$ from the average, where
$\sigma$ is the standard deviation of the yields in a given run range,
are considered outliers. Two such runs were found out of 457
considered. The integrated luminosities of these two runs are 14.3
nb$^{-1}$ and 258.3 nb$^{-1}$.  Further investigations of online
operational conditions during these runs revealed no obvious hardware
or trigger malfunctions. Since the probability is 1\% that a data
subsample of 258.3 nb$^{-1}$ out of a total sample of $39.7$ pb$^{-1}$
would have a yield different by $>4 \sigma$, both runs are included in
the baseline cross-section measurement. The measurement is repeated
without the outlier runs included and a systematic uncertainty
assigned from the difference in the measurements. We find the
uncertainty on the total cross section to be less than $1\%$.

\subsection{Trigger Efficiency}

\begin{figure}[!h]
\centerline{\psfig{figure=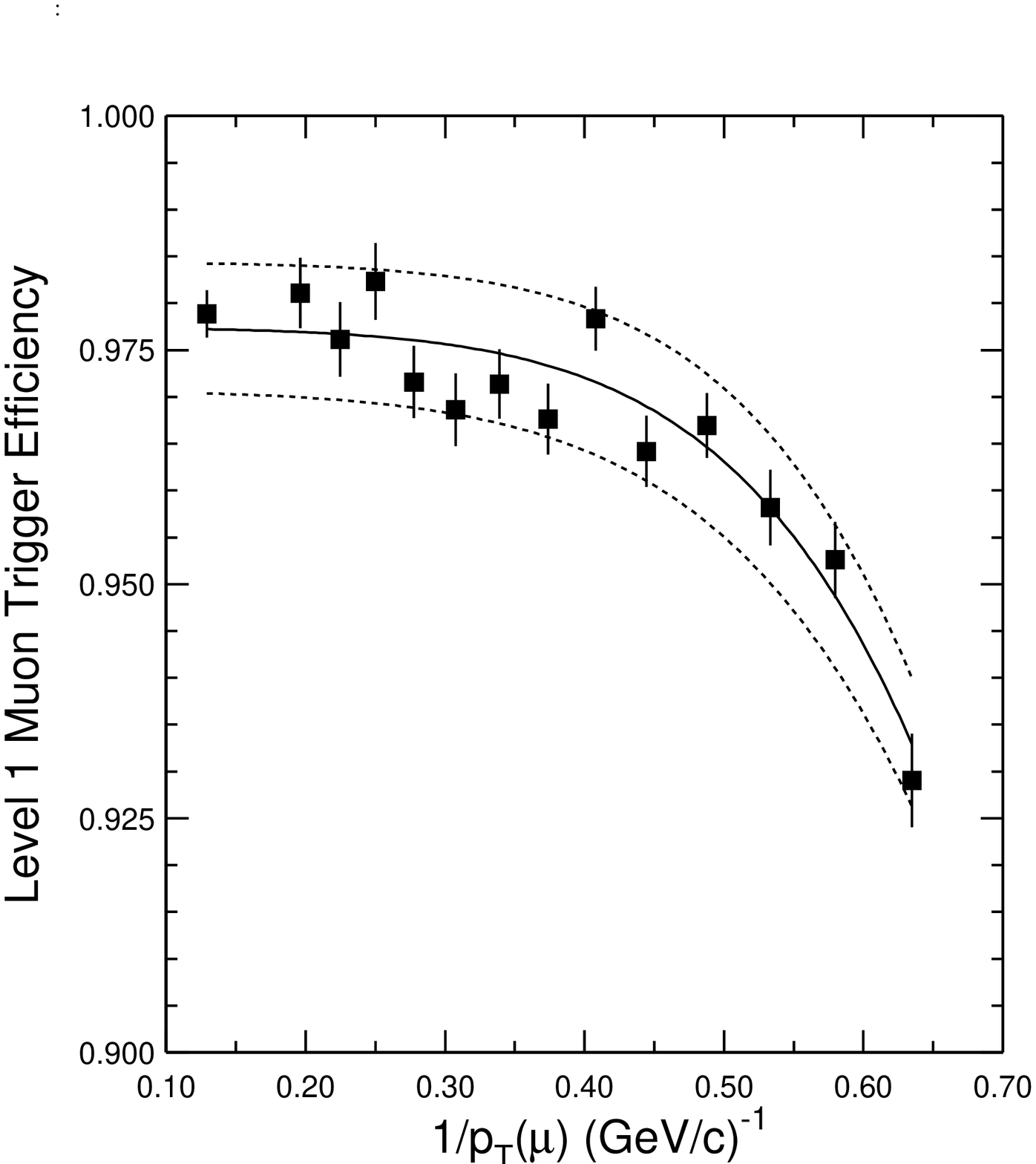,width=0.5\textwidth}
}
\caption{ The Level 1 CMU trigger efficiency as a function of 
muon $p_T$.  Points with error bars are measurement points. The solid line is 
the fitting result using the function described in the text. The 
dashed lines are range used to
determine the uncertainty.
}
\label{fig_trigsys1}
\end{figure}

For our measurement of the Level 1 di-muon trigger efficiency, we used
$J/\psi$ events that were taken with a high-$p_T$ single-muon trigger.  
At Level 1, this trigger requires a muon  with $p_T$ greater than 
$4.0$\,GeV/$c$.
In Level 3, a $J/\psi$ is reconstructed using the triggered 
high-$p_T$ muon 
and a second muon which is not required to pass the Level 1
requirements.  This second  muon  is then used to measure the Level 1
single-muon efficiency.  
The denominator of the efficiency measurement 
is the number of  $J/\psi$ reconstructed using the
Level 3 track and muon information.  
These $J/\psi$ candidates must
have a mass between 2.7 and 3.6~GeV/$c^{2}$, a di-muon opening angle of
$\Delta\phi_0< 130^\circ$, and a separation in $z_0$ 
of less than 5\,cm between the candidate's tracks.  The probe-muon
track must have at least 20 COT axial-layer hits and 16 COT
stereo-layer hits, a CMU $r$-$\phi$ match of $\chi^2(\Delta r\phi) < 9$,
and a track $|z_0| < 90$\,cm. Tracks are excluded if they pass within
1.5\,cm of the center of any of the COT wire planes in
any of the axial layers in order to avoid the inefficient region 
caused by wire supports.  For the probe muon to pass 
the Level 1 trigger,  the associated Level 3 track must be matched
to an XFT track and the Level 3 CMU stub must be matched to a Level 1 
CMU stub that lies within XTRP window.
The resulting Level 1 muon-finding efficiency is shown in 
Fig.~\ref{fig_trigsys1}.
The  distribution is fit to the following function:
\begin{equation}
\epsilon_{\rm L1}^{\mu}(p_T^{\mu}) = E \cdot {\rm freq} \left( \frac{A -
1/p_T}{R} \right),
\label{eqn_l1trig}
\end{equation}
where freq is the normal frequency function:
\begin{equation}
{\rm freq}(x) = \frac{1}{\sqrt{2\pi}}\int_{\rm -\infty}^{x} e^{-\frac{1}{2}
t^2} dt,
\end{equation}
$E$ is the plateau efficiency, $A$ is associated with the $p_T$ at
which the efficiency is half the peak value, and $R$ is the effective
Gaussian resolution.  We find $E = 0.977\pm0.002$, 
$A=1.1\pm0.1$~$(\rm\,GeV/c)^{-1}$,  and $R=0.28\pm0.06$~$(\rm\,GeV/c)^{-1}$.

To determine the uncertainty in the Level 1 trigger efficiency, while also
taking into account the data fluctuations around the central fit as
shown in Fig.~\ref{fig_trigsys1}, the range of the
uppermost and lowermost fluctuations supported by the data are
computed as follows:
$x'(p_T) = \bar{x} \pm ( |x-\bar{x}| + 1\sigma)$
where $x$ is the data value, $\bar{x}$ is the value returned by the
fit and $\sigma$ is the uncertainty on the data. The $x'(p_T)$
distribution is refit using the function in Equation~\ref{eqn_l1trig}. 
The results are shown as dashed lines in Fig.~\ref{fig_trigsys1}.
The di-muon Level 1 trigger efficiency is calculated on an
event-by-event basis to take into account $\mu$-$\mu$ correlations. For
each $J/\psi$ candidate, the Level 1 $J/\psi$ reconstruction
efficiency is given by:
\begin{equation}
\epsilon_{\rm L1}^{J/\psi}(p_T^{J/\psi})
= \epsilon_{\rm L1}^{\mu}(p_T^{\mu_1}) \cdot \epsilon_{L1}^{\mu}(p_T^{\mu_2}),
\end{equation}
where $\epsilon_{\rm L1}^{\mu}(p_T^{\mu})$ is the single muon Level 1
trigger efficiency given by Equation~\ref{eqn_l1trig}, and
$p_T^{\mu_{\rm 1,2}}$ are the transverse momenta of the two muon
candidates.  The trigger's exclusion of pairs with nearby stubs is
included as part of the geometric acceptance.
The mean of the Level 1 di-muon trigger efficiency distribution in
each $J/\psi$ transverse momentum bin is listed in  
Table ~\ref{tab_xsecsum}. The maximum difference  from varying 
the trigger efficiencies by one standard deviation independently 
for the two muons   is listed as the uncertainty on the di-muon trigger  
in Table ~\ref{tab_xsecsum}. We find that the variation is within
$\pm 1.5\%$ in all bins.

The Level 3 reconstruction efficiency is dominated by the difference
between the online and offline tracking efficiency. A fast
tracking algorithm is used for pattern recognition in the
COT in Level 3. In the offline reconstruction a more accurate tracking
algorithm is combined with the result of the Level 3
algorithm to give a higher overall COT tracking
efficiency. The Level 3 single-muon reconstruction efficiency as
measured versus the offline reconstruction
algorithm is found to be constant for $p_T(\mu)>1.5$\,GeV/$c$ and is
\begin{equation}
\epsilon_{\rm L3/Offline}^{\mu} = 0.997 \pm 0.001(stat) \pm 0.002
(syst).
\end{equation}
In the Level 3 trigger, the muons are required to be separated in 
$z_0$ by less than 5\,cm.  The efficiency $\epsilon_{\rm \Delta z_0}$ 
 of this cut is measured using $J/\psi$ candidates reconstructed
in single-muon-trigger data samples where a Level 3 di-muon trigger was
not required to acquire the data. The numbers of events that passed
the $z_0$-separation criterion in the mass signal and sideband regions
are examined. The cut is found to be 100\% efficient with an
uncertainty of 0.1\%. The uncertainty is driven by the statistical
limitations of the small data samples obtained from the single-muon
triggers.

\subsection{Reconstruction Efficiencies}

The COT tracking efficiency was measured using a Monte Carlo track
embedding technique. Hits from simulated muon tracks are
embedded into CDF Run II di-muon events. The distance resolution
and hit-merging distance are adjusted so the embedded track has
residuals and hit distributions matched to muon tracks in
$J/\psi$ data events. The efficiency of COT track reconstruction in
di-muon events is found to be
\begin{equation}
\epsilon_{\rm COT}(p_T^\mu>1.5 {\rm\,GeV/}c) =
0.9961\pm0.0002(stat)^{+0.0034}_{\rm -0.0091}(syst).
\end{equation} 

The absolute offline reconstruction efficiency of muons including stub
reconstruction and matching stubs to tracks is measured using
$J/\psi$ events from single-muon trigger samples where the $J/\psi$ invariant
mass is reconstructed from a triggered, fully-reconstructed muon and a
second track. Tracks from  the di-muon-mass signal region are projected to
the muon chambers,  and the efficiency of finding a matched stub is
measured. For muons in the CMU fiducial 
region with  $p_T(\mu)>1.5$\,GeV/$c$, the
offline reconstruction efficiency is found to be independent of $p_T$ 
and is measured to be:
\begin{equation}
\epsilon_{\rm CMU}^{\mu}= 0.986 \pm 0.003 \pm 0.010.
\end{equation} 

To select clean CMU muons, the track-stub matching in the $r$-$\phi$
plane is required to have $\chi^2(\Delta r\phi) < 9$. The
efficiency of this cut is found to have a weak dependence on $p_T^{\mu}$:
\begin{equation}
 \epsilon_{\rm \chi^2} = (1.0018\pm0.0003) - (0.0024\pm
0.0001) p_T^{\mu}.
\label{eqn_chisq}
\end{equation}
The efficiency of the track-stub matching criterion
($\chi^2(\Delta r\phi) < 9$) as a function of $J/\psi$ transverse
momentum, obtained using an event-by-event weighting is listed in 
Table~\ref{tab_xsecsum}.  The systematic uncertainty on the weighted-average
matching-cut efficiency is obtained by varying the normalization and
slope in Equation~\ref{eqn_chisq} by one standard deviation. The
change in the weighted average efficiency in each $J/\psi$ transverse momentum
bin is found to be $\leq 0.2\%$.

Since the two muons originate from a common decay point, the efficiency of the
track $z_0$ cut is fully correlated for the two muons and is counted only once.
The combined $p_T$ independent COT-tracking, muon and Level 3
reconstruction efficiencies for $J/\psi$ mesons is calculated to be
\begin{equation}
\epsilon_{\rm rec} =  \epsilon_{L3}^2 \cdot \epsilon_{COT}^2 \cdot
\epsilon_{\rm CMU}^2  \cdot \epsilon_{z_0} \cdot
\epsilon_{\rm \Delta_{z_0}} = 95.5 \pm 2.7 \%.
\end{equation}
Table~\ref{tab_effsum} summarizes the $p_T$-independent
reconstruction efficiencies and those of the various muon
selection cuts.
\begin{table*}
\caption{Summary of $J/\psi$ reconstruction efficiencies.}
\begin{tabular}{|cc|}
\hline \hline
$J/\psi$ Selection & Efficiency \\ \hline
Level 3 muon reconstruction  
& $\epsilon_{L3} = 0.997 \pm 0.001 \pm 0.002 $ \\
COT offline tracking    
& $\epsilon_{COT} = 0.9961\pm0.0002^{+0.0034}_{-0.0091}$ \\
Muon offline reconstruction  
& $\epsilon_{CMU}= 0.986 \pm 0.003 \pm 0.010 $\\
Muon $z_0$ position less than $\pm 90$\,cm  
&  $\epsilon_{z_0} = 0.9943 \pm 0.0016$ \\
Di-muon $z_0$ separation less than $5$\,cm 
& $\epsilon_{\Delta z_0} = 1.0 \pm 0.001$ \\
\hline
Total reconstruction 
& $\epsilon_{rec} =  \epsilon_{L3}^2 \cdot \epsilon_{COT}^2 \cdot \epsilon_{CMU}^2  \cdot \epsilon_{z_0} \cdot
\epsilon_{\Delta z_0} = 95.5 \pm 2.7\%$ \\ \hline \hline
\end{tabular}
\label{tab_effsum}
\end{table*}

\section{\boldmath $J/\psi$ Cross Section}

An event-by-event weighting is used to determine the $J/\psi$ yield 
in each $p_T$ bin. Each event is weighted using the Level 1 single muon
efficiency $\epsilon_{\rm L1} (p_T^{\mu})$  and the efficiency of the 
track-stub matching criterion $ \epsilon_{\rm \chi^2} (p_T^{\mu})$ applied
to each of the two muons. The event is then corrected for the 
acceptance  ${\mathcal{A}}(p_T^{J/\psi},y^{J/\psi})$. The weight of
each candidate event is given by:
\begin{eqnarray}
\nonumber
1/w_i  = &&  \epsilon_{\rm L1} (p_T^{\mu1}) \cdot \epsilon_{L1} (p_T^{\mu2}) \cdot \\ 
         &&  \epsilon_{\rm \chi^2}  (p_T^{\mu1}) \cdot \epsilon_{\chi^2} (p_T^{\mu2}) \cdot 
           {\mathcal{A}}(p_T^{J/\psi},y^{J/\psi}). 
\end{eqnarray}
We fit the invariant mass distributions of the weighted events, using
the same shapes for signal and background as 
shown in Fig.~\ref{fig_mass1a}, ~\ref{fig_mass1d},~\ref{fig_mass1b} 
and \ref{fig_mass1c}. The number
of signal events in each transverse momentum bin is determined from
the area under the signal mass peak.  The error on the corrected yield from
the mass template fit, $N(p_T)_{\rm corrected}$ is given by:
\begin{equation}
\delta (N(p_T)_{\rm corrected}) = \sqrt{ \sum_{i=0}^{i=N_s} \left( {w_i}
\right)^2},
\end{equation}
where $N_s$ is the raw number of signal events in each momentum bin
before weighting. In a similar fashion, the di-muon $p_T$ distribution
in each bin is weighted. The weighed $p_T$ distribution of the mass
sideband subtracted events in the $J/\psi$ mass signal region is used to
determine the mean $p_T$ value for each transverse momentum bin.

The $J/\psi$ differential cross section is then calculated as follows:
\begin{equation}
{ d\sigma   
\over dp_T} \cdot Br(J/\psi \rightarrow \mu \mu) 
= {N(p_T)_{\rm corrected}  \cdot  (1-{\mathcal{A}'}) \over
\epsilon_{\rm rec} \cdot 
\int{ {\cal L}} dt  
\cdot \Delta p_T},
\end{equation}
where $d\sigma/dp_T$ is the average cross section of inclusive $J/\psi$ in
that $p_T$ bin integrated over $\mid y (J/\psi) \mid < 0.6$,
${\mathcal{A}'}$ is the correction factor for $y$ smearing defined by
Equation~\ref{eqn_ysmear},
$\epsilon_{\rm rec}$ is the combined Level 3 and offline tracking and muon
reconstruction efficiency, $\int{ {\cal L} dt}$ is the integrated
luminosity, and $\Delta p_T$ is the size of the $p_T$ bin.

The cross-section values obtained with statistical and $p_T$-dependent
uncertainties are listed
in Table~\ref{tab_xsec}.

\begingroup
\squeezetable
\begin{table*}
\caption{
 The differential $J/\psi$ cross section times  the
branching fraction $Br \equiv Br(J/\psi \rightarrow \mu \mu)$ 
as a function of $p_T$ for $|y(J/\psi)|<0.6$.  For each measurement, 
the first uncertainty is
statistical and the second uncertainty is systematic. The
systematic uncertainties shown are the $p_T$ dependent uncertainties
only. The fully correlated $p_T$ independent systematic uncertainty in
each bin is $7\%$.}
\begin{center}
\begin{tabular}{ccccc} 
\hline \hline
$p_T(J/\psi)$  (\,GeV/$c$) & Mean $p_T$  & Mean $p_T^2$ & $\frac{d\sigma}{dp_T}
\cdot Br$ (nb/(\,GeV/$c$)) &  $\frac{d\sigma}{dp_T^2} \cdot Br$ (nb/(\,GeV/{\it{c}}$^{2}$))  \\  \hline    		      
$0.0-0.25$  & $ 0.15 $&$0.027$&$ 9.13 \pm 0.6 { (stat)} ^{+1.1}_{-0.7}
{ (syst)}    $&$36.5 \pm 2.4 { (stat)} ^{+4.2}_{-2.6} { (syst)} $\\
$0.25-0.5$  & $ 0.39 $&$0.16 $&$ 28.1 \pm 1.5  ^{+2.4}_{-1.6}    $&$37.4 \pm 2.0^{+3.1}_{-2.0} $\\
$0.5-0.75$  & $ 0.64 $&$0.42 $&$ 45.3 \pm 1.9  ^{+3.0}_{-2.1}    $&$36.2 \pm 1.5^{+2.5}_{-1.8} $\\
$0.75-1.0$  & $ 0.89 $&$0.79 $&$ 59.3 \pm 2.0  ^{+4.0}_{-2.9}    $&$33.9 \pm 1.1^{+2.3}_{-1.6} $\\
$1.0-1.25$  & $ 1.13 $&$1.29 $&$ 69.6 \pm 1.9  ^{+3.6}_{-3.2}    $&$31.0 \pm 0.8^{+1.7}_{-1.5} $\\
$1.25-1.5$  & $ 1.38 $&$1.91 $&$ 73.4 \pm 1.7  ^{+3.9}_{-3.5}    $&$26.7 \pm 0.6^{+1.4}_{-1.3} $\\
$1.5-1.75$  & $ 1.63 $&$2.66 $&$ 75.2 \pm 1.6  ^{+3.8}_{-3.3}    $&$23.2 \pm 0.5^{+1.2}_{-1.0} $\\
$1.75-2.0$  & $ 1.87 $&$3.52 $&$ 72.9 \pm 1.4  ^{+3.7 }_{-3.3}   $&$19.4 \pm 0.4^{+0.9}_{-0.8} $\\
$2.0-2.25$  & $ 2.13 $&$4.53 $&$ 69.1 \pm 0.8  ^{+3.3 }_{-2.9}   $&$16.3 \pm 0.2^{+0.8}_{-0.7} $\\
$2.25-2.5$  & $ 2.38 $&$5.65 $&$ 67.3 \pm 1.0  ^{+3.1 }_{-2.8}   $&$14.2 \pm 0.2^{+0.7}_{-0.6} $\\
$2.5-2.75$  & $ 2.62 $&$6.89 $&$ 57.6 \pm 0.9  \pm 2.6   $&$11.0 \pm 0.2 \pm 0.5  $\\
$2.75-3.0$  & $ 2.87 $&$8.26 $&$ 52.0 \pm 0.8  \pm 2.4   $&$9.04 \pm 0.13 \pm 0.41 $\\
$3.0-3.25$  & $ 3.12 $&$9.76 $&$ 43.6 \pm 0.7  \pm 1.9   $&$6.97 \pm 0.10 \pm 0.31 $\\
$3.25-3.5$  & $ 3.38 $&$11.4 $&$ 37.3 \pm 0.6  \pm 1.6   $&$5.53 \pm 0.08 \pm 0.24 $\\
$3.5-3.75$  & $ 3.62 $&$13.1 $&$ 31.5 \pm 0.5  \pm 1.3   $&$4.34 \pm 0.07 \pm 0.18 $\\
$3.75-4.0$  & $ 3.87 $&$15.0 $&$ 26.2 \pm 0.4  \pm 1.2   $&$3.38 \pm 0.05 \pm 0.15 $\\
$4.0-4.25$  & $ 4.12 $&$17.0 $&$ 22.5 \pm 0.4  \pm 1.0   $&$2.72 \pm 0.05 \pm 0.12 $\\
$4.25-4.5$  & $ 4.38 $&$19.2 $&$ 18.7 \pm 0.3  \pm 0.8   $&$2.13 \pm 0.04 \pm 0.09 $\\
$4.5-4.75$  & $ 4.62 $&$21.4 $&$ 16.1 \pm 0.3  \pm 0.7   $&$1.74 \pm 0.03 \pm 0.08 $\\
$4.75-5.0$  & $ 4.88 $&$23.8 $&$ 13.3 \pm 0.3  \pm 0.6   $&$1.37 \pm 0.03 \pm 0.06 $\\
$5.0-5.5$   & $ 5.24 $&$27.5 $&$ 10.3 \pm 0.15 \pm 0.42  $&$0.984 \pm 0.014 \pm 0.040 $\\
$5.5-6.0$   & $ 5.74 $&$33.0 $&$ 7.28 \pm 0.12 \pm 0.29  $&$0.633 \pm 0.010 \pm 0.025 $\\
$6.0-6.5$   & $ 6.24 $&$38.9 $&$ 5.11 \pm 0.09 \pm 0.20  $&$0.408 \pm 0.0069 \pm 0.016 $\\
$6.5-7.0$   & $ 6.74 $&$45.5 $&$ 3.54 \pm 0.07 \pm 0.14  $&$0.262 \pm 0.0052 \pm 0.010 $\\
$7.0-8.0$   & $ 7.45 $&$55.7 $&$ 2.27 \pm 0.03 \pm 0.10  $&$0.151 \pm 0.0019 \pm 0.006 $\\
$8.0-9.0$   & $ 8.46 $&$71.6 $&$ 1.14 \pm 0.02 \pm 0.05  $&$0.0668 \pm 0.0011 \pm 0.0028 $\\ 
$9.0-10.0$  & $ 9.46 $&$89.5 $&$0.622 \pm 0.013 \pm 0.025 $&$0.0327 \pm 0.0007 \pm 0.0013 $\\ 
$10.0-12.0$ & $ 10.8 $&$118 $&$0.278 \pm 0.006 \pm 0.011 $&$0.0126 \pm 0.0003 \pm 0.0005 $\\  
$12.0-14.0$ & $ 12.8 $&$165 $&$0.103 \pm 0.003 \pm 0.004 $&$0.00398 \pm 0.00013 \pm 0.00015 $\\
$14.0-17.0$ & $ 15.2 $&$233  $&$0.037 \pm 0.002 \pm 0.002 $&$0.00120 \pm 0.00005 \pm 0.00006 $\\
$17.0-20.0$   & $ 18.3 $&$336  $&$0.014 \pm 0.001 \pm 0.001 $&$0.00037 \pm 0.00004 \pm 0.00002 $\\
\hline \hline
\end{tabular}
\end{center}
\label{tab_xsec}
\end{table*}
\endgroup

An uncertainty of +0.1$\%$ on the momentum scale is extracted by 
comparing the reconstructed $J/\psi$ mass as shown in 
Fig.~\ref{jpsi-rawmass.eps} to the world averaged value
of $3.09688 \pm 0.00004$\,GeV/$c^2$~\cite{PDG}.  
The  3 MeV/$c^2$  difference is attributed to an underestimation of 
the energy loss in the silicon detector due to an incomplete 
accounting of the material at the time the data sample used in this analysis 
was processed.
The $+0.1\%$ uncertainty from the momentum
scale corresponds to an uncertainty on the differential cross section 
as  $  d(d\sigma/dp_T)/dp_T   \times 0.1\%$. 
Using the values  in Table~\ref{tab_xsec}, the first derivative of the
differential cross section is calculated and the momentum scale
uncertainty on the cross section in each bin estimated. The effect was
found to be small, the largest negative deviation being $-0.08\%$ and
the largest positive deviation being $+0.7\%$.

Table~\ref{tab_syst} summarizes the different contributions to the
systematic errors  applied to the cross-section measurement from
acceptance calculations using a Monte Carlo simulation, the mass line
shapes used to determine the yield, the trigger and reconstruction
efficiencies, and the luminosity measurement.

\begin{table}[!ht]
\caption{Summary of systematic uncertainties in the inclusive $J/\psi$
cross-section measurement.  The $p_T$ dependent uncertainties are listed in 
the top section of the table. In general, the $p_T$ dependent
uncertainties increase with decreasing $p_T$. The total is calculated from 
the $p_T$ independent sources only.
}
\begin{center}
\begin{tabular}{|ccc|} 
\hline \hline
& Source & Size \\ \hline 
Acceptance & $ J/\psi$ spin alignment &$\pm (2\rightarrow5)\%$ ($p_T$)	\\
Acceptance & $p_T$ spectrum            & $\pm (0\rightarrow5)\%$ ($p_T$)\\
Acceptance & Detector  material        & $\pm (0.4\rightarrow 5)\%$ ($p_T$)\\
Yield      & Mass fits       & $ (-1.3  \rightarrow +9)\%$ ($p_T$) \\
Yield      & Momentum scale  & $ (-0.1 \rightarrow  +0.7)\%$ ($p_T$) \\ \hline 
Luminosity & CLC             & $\pm 6.0\%$ \\
Reconstruction & Table~\ref{tab_effsum} & $\pm 2.8\%$ \\
Acceptance & CMU simulation           & $\pm 1.0\%$ \\
Yield 	   & Data quality                  & $\pm 1.0\%$ \\
L1 trigger efficiency  &  Table~\ref{tab_xsecsum} & $\pm 1.5\%$ \\ \hline 
Total  & \multicolumn{2}{c|}{ $\pm 6.9\% \oplus \delta(p_T)$} 
  \\ \hline \hline
\end{tabular}
\end{center}
\label{tab_syst}
\end{table}

\begin{figure}[!h]
\centerline{\psfig{figure=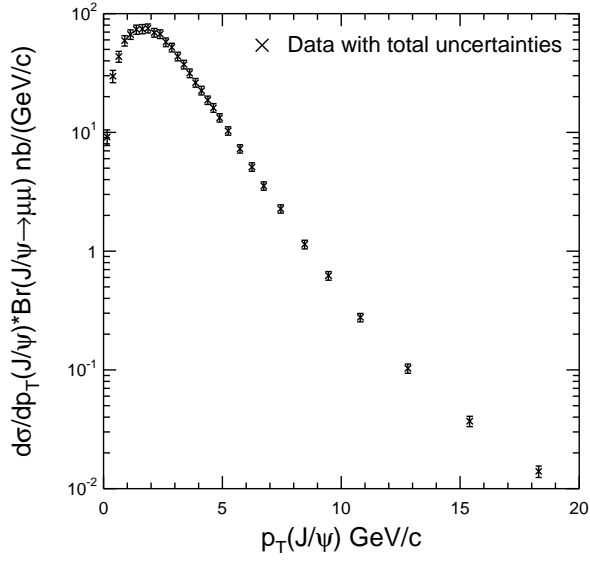,width=0.5\textwidth}
}
\caption{Inclusive $J/\psi$ cross section, 
$d\sigma/dp_T \cdot Br(J/\psi \rightarrow \mu \mu)$,
 as a function of $J/\psi$
$p_T$ integrated over the rapidity range $|y|<0.6$.  The differential
cross section with systematic and statistical uncertainties added is
plotted. This includes correlated uncertainties. }
\label{Jpsi_xsec_syst.eps}
\end{figure}

\begin{figure}[!h]
\centerline{\psfig{figure=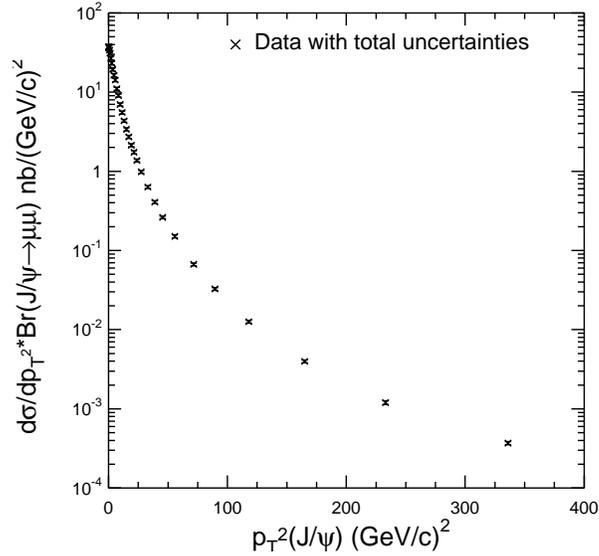,width=0.5\textwidth}
}
\caption{The invariant  $J/\psi$ cross section, 
$d\sigma/dp_T^2 \cdot Br(J/\psi \rightarrow \mu \mu)$,
 as a function of  $p_T^2 (J/\psi)$.
The differential cross section  
is plotted. This includes correlated uncertainties. }
\label{fig_ptsqr}
\end{figure}

The differential cross-section results with systematic and statistical
uncertainties are displayed in Fig.~\ref{Jpsi_xsec_syst.eps}. 
The invariant cross section,
$d\sigma/dp_T^2 \cdot Br(J/\psi \rightarrow \mu \mu) $, with systematic errors is shown in Fig.~\ref{fig_ptsqr}. 
The results are also listed
in Table~\ref{tab_xsec}.

We integrate the 
differential cross section to find the total
$J/\psi$ production cross section:
\begin{eqnarray}
\nonumber
&& \sigma(p \overline{p} \rightarrow J/\psi X,\mid y (J/\psi) \mid <
0.6) \cdot Br(J/\psi \rightarrow \mu \mu) \\
&& = \TOTXSECBR \ {\rm nb}.
\end{eqnarray}
The $p_T$-dependent systematic uncertainties are summed and then
added in quadrature with the fully-correlated uncertainty of 6.9\%:
\begin{eqnarray}
\delta_{\rm tot \ \sigma \cdot {rm Br} }^{stat} &=& \sqrt{\sum_{i=1}^{N_{\rm bins}}
 (\delta_i^{stat})^2} = 1 \ {\rm nb} \\ \nonumber
\delta_{\rm tot \ \sigma \cdot {rm Br} }^{syst} &=& \left\{ \sum_{i=1}^{N_{\rm bins}}
 \delta_i^{syst} (p_T) \right\} \oplus \pm 16 \ {\rm nb} \\ 
& = & ^{+21}_{\rm -19} \ {\rm nb},
\end{eqnarray}
where $N_{\rm bins}$ is the total number of $p_T$ bins,
$\delta_i^{stat}$ is the statistical uncertainty in the cross-section
measurement in the $i^{th}$ bin, $\delta_i^{syst} (p_T)$ is the
systematic uncertainty on the measurement in each $p_T$ bin
independent of the correlated systematic uncertainty of 6.9\%, and
$\oplus$ denotes addition in quadrature.  After correcting for the
$Br(J/\psi
\rightarrow \mu \mu) = 5.88 \pm 0.10 \%$~\cite{PDG}, we find
\begin{eqnarray}
\nonumber
&& \sigma(p \overline{p} \rightarrow J/\psi X,\mid y (J/\psi) \mid < 0.6) \\
&& = \TOTXSEC \ \mu{\rm b}.
\end{eqnarray}

To compare with prior measurements where only the portion of the cross
section for  $p_T(J/\psi)$ exceeding 5\,GeV/$c$ was 
measured~\cite{RUNIPSI}, we also measure the integrated cross section of
inclusive $J/\psi$ with $p_T>5$\,GeV/$c$ and $|\eta|<0.6$ at
$\sqrt{s}=1960$\,GeV. We find the cross section is
\begin{eqnarray}
\nonumber 
&& \sigma( p \overline{p} \rightarrow J/\psi X, 
p_T (J/\psi)>5.0 \ {\rm\,GeV/}c, \mid \eta (J/\psi) \mid <
0.6) \cdot Br(J/\psi \rightarrow \mu \mu) \\ 
&& = \XSECGTFIVEETA \ {\rm nb}. 
\label{eqn_incgt5}
\end{eqnarray}
We discuss the comparison of this result with earlier data in
Section~\ref{sec_discuss}.

\section{\boldmath $H_b \rightarrow J/\psi$ Fraction and the $b$-hadron Cross Section}
In general, the inclusive $J/\psi$ cross section contains
contributions from various sources: prompt production of charmonium; 
decays of excited charmonium states such as $\psi(2S),
\ \chi_{\rm c1}$ and $\chi_{c2}$; and decays of $b$-hadrons. The
charmonium states decay immediately. In contrast, $b$-hadrons have
long lifetimes that are on the order of picoseconds.  
This implies that $J/\psi$ events from the decays of $b$-hadrons are 
likely to be displaced from the beamline. 
We exploit this feature to separate $J/\psi$ of decay products of 
$b$-hadrons from that of prompt charmonium in the $p_T$ bins used in the 
inclusive $J/\psi$ cross section calculation.

To measure the fraction of $J/\psi$ events that are from displaced decay 
vertices, we use the subset of the $J/\psi$ sample that includes  
those events for which both muon tracks from the
$J/\psi$ satisfy high quality COT-SVX II track requirements. 
The track
extrapolation from the path formed by the trajectory in the COT into
the SVX II is described in Section~\ref{sec_track}. The total number
of hits expected in the five layers of the SVX II is determined from
the number of functioning and powered silicon sensors intersected by
the COT muon track. Tracks missing more than one expected hit in the
SVX II are rejected. Both tracks are required to have a hit in the
innermost layer of the SVX II and a hit in the second layer if the
sensor intersected by the COT track is functioning.
Corrections for energy loss in the SVX II are applied to the candidate
muons based on a GEANT simulation of the material.  The two muon
tracks are constrained to come from a common space point.  The
$\chi^2$ probability of this 3-dimensional vertex fit is required 
to exceed  0.001. We find that $139200 \pm 500$ events, or about half of the
total $J/\psi$ data sample, pass these criteria. While the data sample
is reduced by the SVX II requirements, the momentum, angle, and vertex
resolutions are substantially improved.

The primary vertex,  taken as the beam position in the $r$-$\phi$ plane,  
is assumed as the point where $b$-hadrons are produced. 
It is  calculated on a run-by-run basis from a data sample  taken using 
the inclusive jet trigger which has 
negligible contributions from charm and
bottom decays so  the beamline position can be calculated with no bias
from detached decay vertices. 
The resolution of the primary vertex in the $r$-$\phi$ plane is limited 
by the $\sim 30~\mu$m  RMS spread in the size of the beam envelope.

\subsection{\boldmath Measurement of the Fraction of $J/\psi$ Events from $b$-hadrons}

The $J/\psi$ from the decay of $H_b \rightarrow J/\psi X $  is likely to be 
displaced from the primary vertex where $b$-hadrons are assumed to be 
produced.   
The signed projection of the flight distance of $J/\psi$ on its 
transverse momentum, $L_{\rm xy}$, is a good measurement of the 
displaced  vertex and can be used  as  a variable to separate 
$J/\psi$ of the  $H_b$ decay products from that of prompt decays. 
This method works well for events with high $J/\psi$ $p_T$ where the 
flight direction aligns well with that of the $b$-hadron.  
For events with  very low $J/\psi$ $p_T$, the non-negligible 
amount of $J/\psi$ with large opening angle between its flight direction 
and that of the $b$-hadron will impair the separation ability.  
Monte Carlo simulation shows that a reliable $b$-fraction can be extracted 
using this method for events with $J/\psi$ $p_T$ greater than 1.25\,GeV/$c$.

The $L_{\rm xy}$ is calculated as
\begin{equation}
L_{\rm xy}(J/\psi) = \vec{L} \cdot \vec{p}_T(J/\psi) / |p_T(J/\psi)|,
\end{equation}
where $\vec{L}$ is the vector from the primary vertex to the $J/\psi$
decay vertex in the $r$-$\phi$ plane and $\vec{p}_T(J/\psi)$ is the
transverse momentum vector. 
To reduce the dependence on the $J/\psi$ transverse momentum bin size
and placement, a new variable $x$, called pseudo-proper decay time,
 is used instead of $L_{\rm xy}$,  
\begin{equation}
x=L_{\rm xy}(J/\psi)\cdot M(J/\psi)/p_T(J/\psi), 
\label{eqn_xdef}
\end{equation}
where the  $M(J/\psi)$ is taken as the known $J/\psi$ mass~\cite{PDG}.  
A Monte Carlo simulation is needed to model the
distribution of $x(J/\psi)$ from $b$-hadron events. The Monte Carlo
templates of the $x$ distributions 
$\mathcal{X}_{\rm mc}(x,p_T^{\rm J/\psi})$ are generated for all $J/\psi$
transverse momentum ranges and are directly convoluted with the value
of the $x$ resolution function measured in the data without allowing
any of the parameters governing the shape of the Monte Carlo
distributions to vary.

\subsubsection{The Likelihood Function}

An unbinned maximum likelihood fit is used to extract the
$b$-fraction, $f_B$, from the data.  The $J/\psi$ pseudo-proper decay time $x$,
its error $\sigma$, and the  mass of the di-muon pair $m_{\rm\mu\mu}$ are 
the input variables. A simultaneous mass and lifetime
fit is performed using a log-likelihood function ($\ln
\mathcal{L}$) given by:
\begin{equation}
\ln {\cal L} = \sum_{\rm i=1}^{N} \ln {\cal F}(x,m_{\mu\mu}),
\end{equation} 
where $N$ is the total number of events in the mass range
$2.85<m_{\rm \mu\mu}<3.35 $\,GeV/$c^2$.

The mass and pseudo-proper decay time distribution is described
by the following function,
\begin{eqnarray}
\nonumber
{\cal F}(x,m_{\rm \mu\mu}) &=& f_{\rm Sig}\times
{\cal F}_{\rm Sig} (x) \times {\cal M}_{\rm Sig}(m_{\mu\mu}) \\
&+& (1-f_{\rm Sig})
\times {\cal F}_{\rm Bkg}(x)
\times {\cal M}_{\rm Bkg}(m_{\mu\mu}),  
\end{eqnarray} 
where  $f_{\rm Sig}$ is the fraction of signal $J/\psi$ events in the mass
region,  ${\cal F}_{Sig}$ and ${\cal F}_{Bkg}$ are the
functional forms describing the $J/\psi$ pseudo-proper decay time distribution
for the signal and background events respectively, 
and ${\cal M}_{\rm Sig}$ and  ${\cal M}_{\rm Bkg}$
 are the functional forms describing the invariant mass
distributions for the signal and background events respectively. We now
describe these components of the likelihood fit in more detail.


The function for modeling the $J/\psi$ pseudo-proper decay time
signal distribution consists of two parts, the $H_b\rightarrow J/\psi X$ decay
and prompt decay functions labeled ${\cal F}_B(x)$ and ${\cal F}_P(x)$
respectively:
\begin{equation}
{\cal F}_{\rm Sig}(x) = \left[f_B \cdot {\cal F}_B(x) + (1-f_B)
\cdot {\cal F}_P(x)\right], 
\end{equation}
where $f_B$ is the fraction of $J/\psi$ mesons originating 
in $b$-hadron decays. We use the $x$ distributions ${\mathcal{X}}_{\rm mc}$ of
accepted events from a Monte Carlo simulation as templates for the $x$
distribution of $b$-hadron events in data. The generated distributions
are convoluted with a resolution function $ R (x'-x, s\sigma)$ such
that the $H_b
\rightarrow J/\psi X $ signal shape is given by
\begin{equation}
{\cal F}_B(x) = R (x'-x, s\sigma) \otimes {\mathcal{X}}_{\rm mc}(x'),
\end{equation}
where $s$ is an overall error scale factor which represents the
possible errors in determining the lifetime resolution and $\otimes$
denotes a convolution. Prompt  $J/\psi$ mesons are produced at the primary vertex, 
therefore their observed displacement
is described only by the resolution function ${\cal F}_P =
R(x,s\sigma)$. We find that $ R (x'-x, s\sigma)$ is best described by
a sum of two Gaussian distributions centered at $x=0$.

The background requires
 a more complicated
parameterization to obtain a good fit to the data outside the $J/\psi$
signal region. The pseudo-proper decay time background function is composed of
four parts: the zero lifetime component, a positive slope exponential
function, a negative slope exponential function, and a symmetric
exponential function with both positive and negative slopes. The
positive slope exponential function is chosen to model the background
from other long lived $b$-hadron events that produce opposite sign
muons such as $b\rightarrow c \mu^- \bar{\nu} X, c \rightarrow \mu^+ \nu
X$. The zero lifetime component is chosen to be the same shape as the
resolution function. The symmetric and negative slope exponential
functions are added to parameterize the remaining components of the
background pseudo-proper decay time distributions which are from unknown
sources. The background exponential tails are also convoluted with the
resolution function.

The background functional form is parameterized as follows:
\begin{eqnarray} \nonumber
&{\cal F}_{\rm Bkg}(x)=&(1-f_{+}-f_{-} -f_{\rm sym}) R(x,s \sigma)\\ \nonumber 
&& +{f_{\rm +} \over \lambda_{+}} \exp(-{x' \over \lambda_{+}}) \theta(x')  
\otimes R (x'-x, s\sigma)\\ \nonumber 
&& +{f_{\rm -} \over \lambda_{-}} \exp({x' \over \lambda_{-}}) \theta(-x')
\otimes R (x'-x, s\sigma)  \\ \nonumber
&& +{f_{\rm sym} \over 2\lambda_{sym}} \exp(-{x'\over\lambda_{sym}}) \theta(x')
\otimes R (x'-x, s\sigma)  \\ 
&& +{f_{\rm sym} \over 2\lambda_{sym}}\exp({x'\over \lambda_{sym}}) \theta(-x')
\otimes R (x'-x, s\sigma),
\end{eqnarray}
where $f_{\rm \pm,~\rm{sym}}$ is the fraction of the background distribution
in the positive, negative and symmetric exponential tails
respectively, $\lambda_{\rm \pm,sym}$ are the corresponding
exponential slopes, and $\theta (x)$ is the step function defined as 
$\theta (x) = 1 $ for $x \geq 0$ and $\theta (x) = 0$ for $x < 0$.  It should 
be kept in mind that the background 
strongly depends on $p_T$ and $m_{\rm \mu\mu}$, and that the likelihood
function incorporates a global fit over the full mass window shown in
Fig.~\ref{fig_mass1a} to Fig.~\ref{fig_mass1c}, including the $J/\psi$
peak and mass sidebands.

The mass resolution used in the
likelihood fit is better than that shown in Figs.~3 - 5 because
of the addition of SVX II hits to the tracks. For the likelihood fit,
the di-muon mass shape ${\cal M}_{\rm Sig}$ is chosen to be simply 
the sum of two Gaussian distributions.
The means of the Gaussian distributions are allowed to
float independently:
\begin{eqnarray}
\nonumber
{\cal M}_{\rm Sig}(m_{\mu\mu}) & = & G_1(m_{\mu\mu}-M, \sigma_M) \\  
&+& f_2 \cdot G_2(m_{\rm \mu\mu}-(M+D), r_2 \sigma_M). 
\end{eqnarray}
The mass fit parameters are  the mean $M$ of the mass
distribution, the width $\sigma_M$ of the first Gaussian distribution,  
the fraction  $f_2$ of the second Gaussian distribution,  the shift $D$ 
in the mean of the
second Gaussian distribution, and the ratio $r_2$ of the widths of the two
Gaussian distributions. 
The mass background is modeled using a linear distribution. This fit 
is adequate for the SVX~II
constrained di-muon mass. The function used, normalized to unity over
the mass range $m^{min}$ to $m^{max}$ is:
\begin{eqnarray}
\nonumber
{\cal M}_{\rm Bkg}(m_{\mu\mu}) & = &
\frac{1}{m_{\rm \mu\mu}^{max}-m_{\mu\mu}^{min}} \\ 
&+& M_{\rm slope} (m_{\mu\mu} - \frac{m_{\mu\mu}^{max}+m_{\mu\mu}^{min}}{2}), 
\end{eqnarray}
where $M_{\rm slope}$ is the slope of the mass background
distribution. The only fit parameter is $M_{\rm slope}$.

 
\subsubsection{The Fits and Systematic Uncertainties}

\begin{figure}[!h]
\centerline{
\psfig{figure=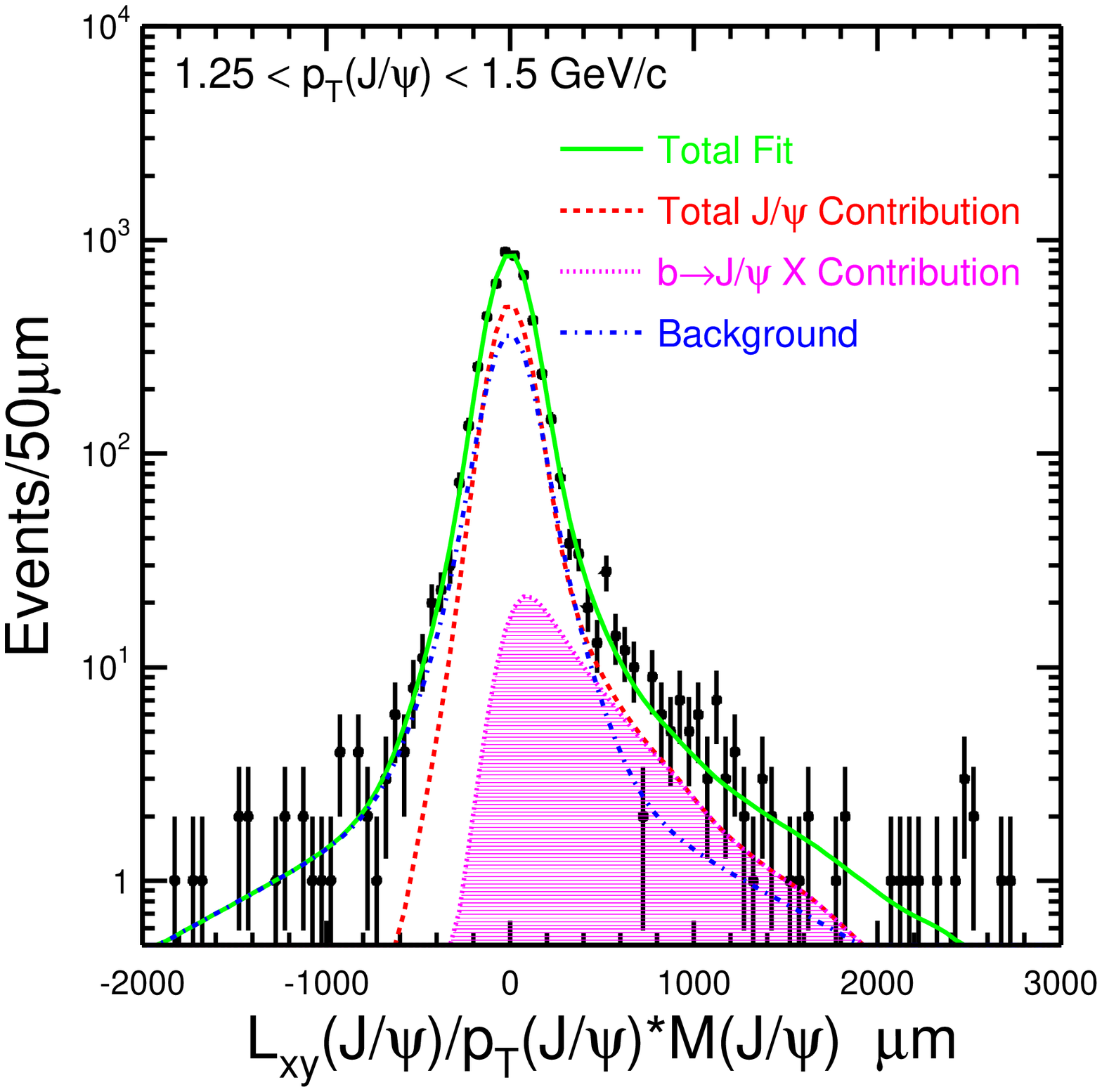,width=0.5\textwidth}
}
\caption{Fits to the $J/\psi$ pseudo-proper decay time in the range
$1.25<p_T(\mu\mu)<1.5$\,GeV/$c$ to extract the fraction of events from
long-lived $b$-hadron decays. The solid line is the fit to all the
events in the mass window of 2.85 to 3.35\,GeV/$c^2$, the dashed line
is the fit to all signal events, the solid histogram is the fit to the
portion of the signal events that are from $b$-hadron decays and the
dot-dashed line is the fit to background events including events in
the invariant mass sidebands.}
\label{fig_bfracfita}
\end{figure}
\begin{figure}[!h]
\centerline{
\psfig{figure=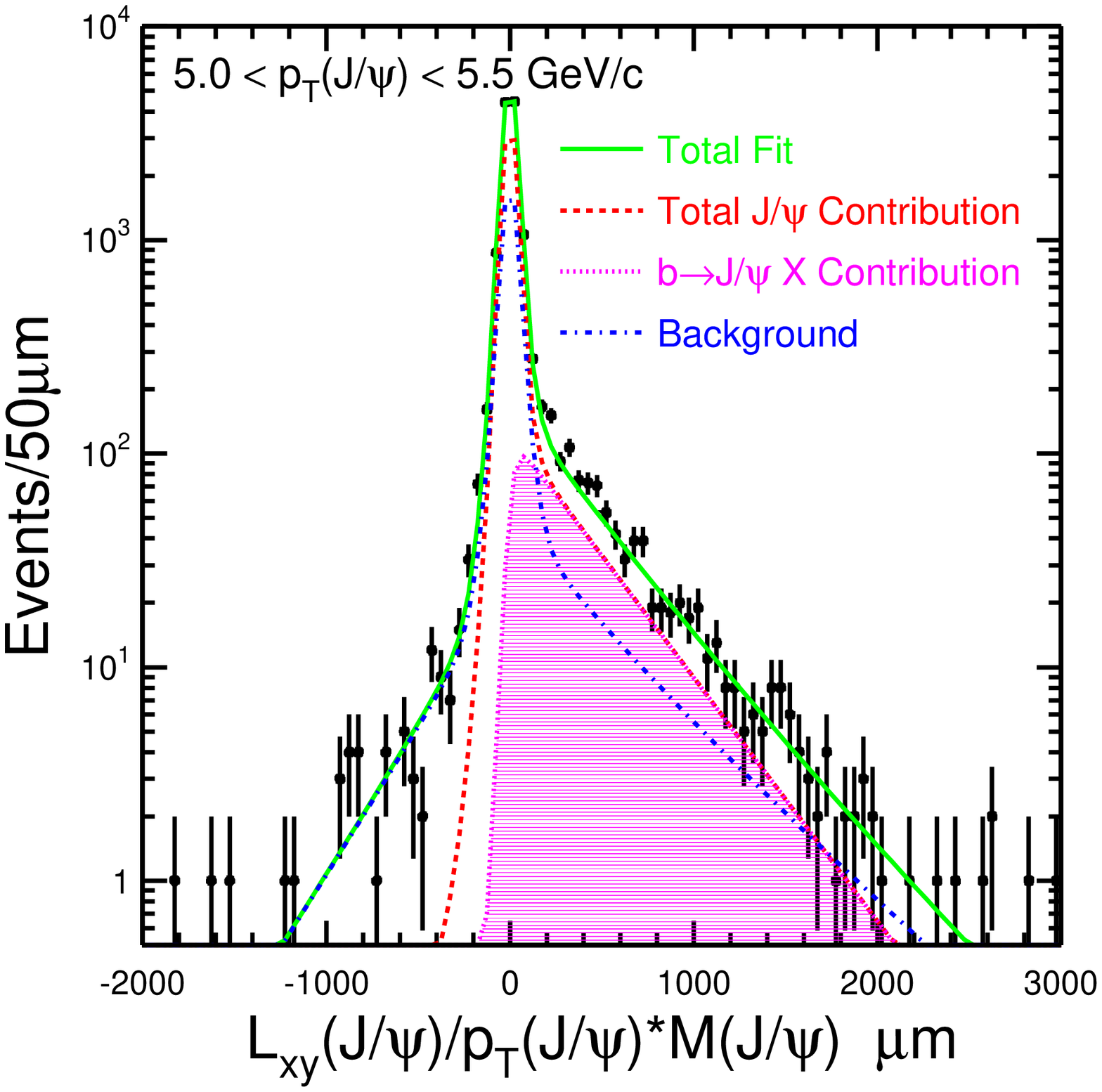,width=0.5\textwidth}
}
\caption{Fits to the $J/\psi$ pseudo-proper decay time in the range
$5.0<p_T(\mu\mu)<5.5$\,GeV/$c$ to extract the fraction of events from
long-lived $b$-hadron decays. The solid line is the fit to all the
events in the mass window, the dashed line is the fit to all signal
events, the solid histogram is the fit to the portion of the signal
events that are from $b$-hadron decays and the dot-dashed line is the
fit to background events. }
\label{fig_bfracfitb}
\end{figure}
\begin{figure}[!h]
\centerline{
\psfig{figure=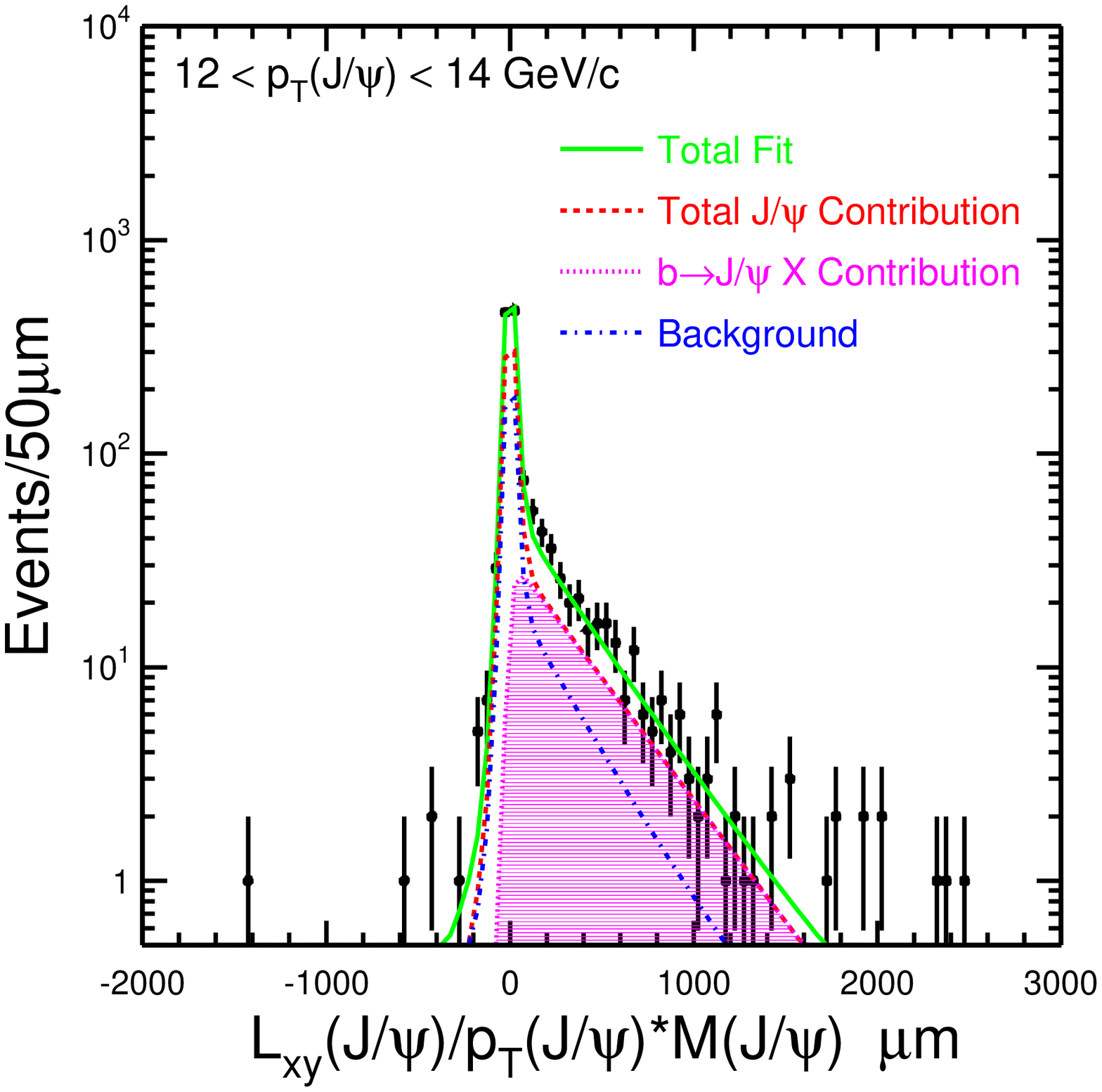,width=0.5\textwidth}
}
\caption{Fits to the $J/\psi$ pseudo-proper decay time  in the range
$12.0<p_T(\mu\mu)<14.0$\,GeV/$c$ to extract the fraction of events from
long-lived $b$-hadron decays. The solid line is the fit to all the
events in the mass window, the dashed line is the fit to all signal
events, the solid histogram is the fit to the portion of the signal
events that are from $b$-hadron decays and the dot-dashed line is the
fit to background events. }
\label{fig_bfracfitc}
\end{figure}

\begin{figure}[!h]
\centerline{\psfig{figure=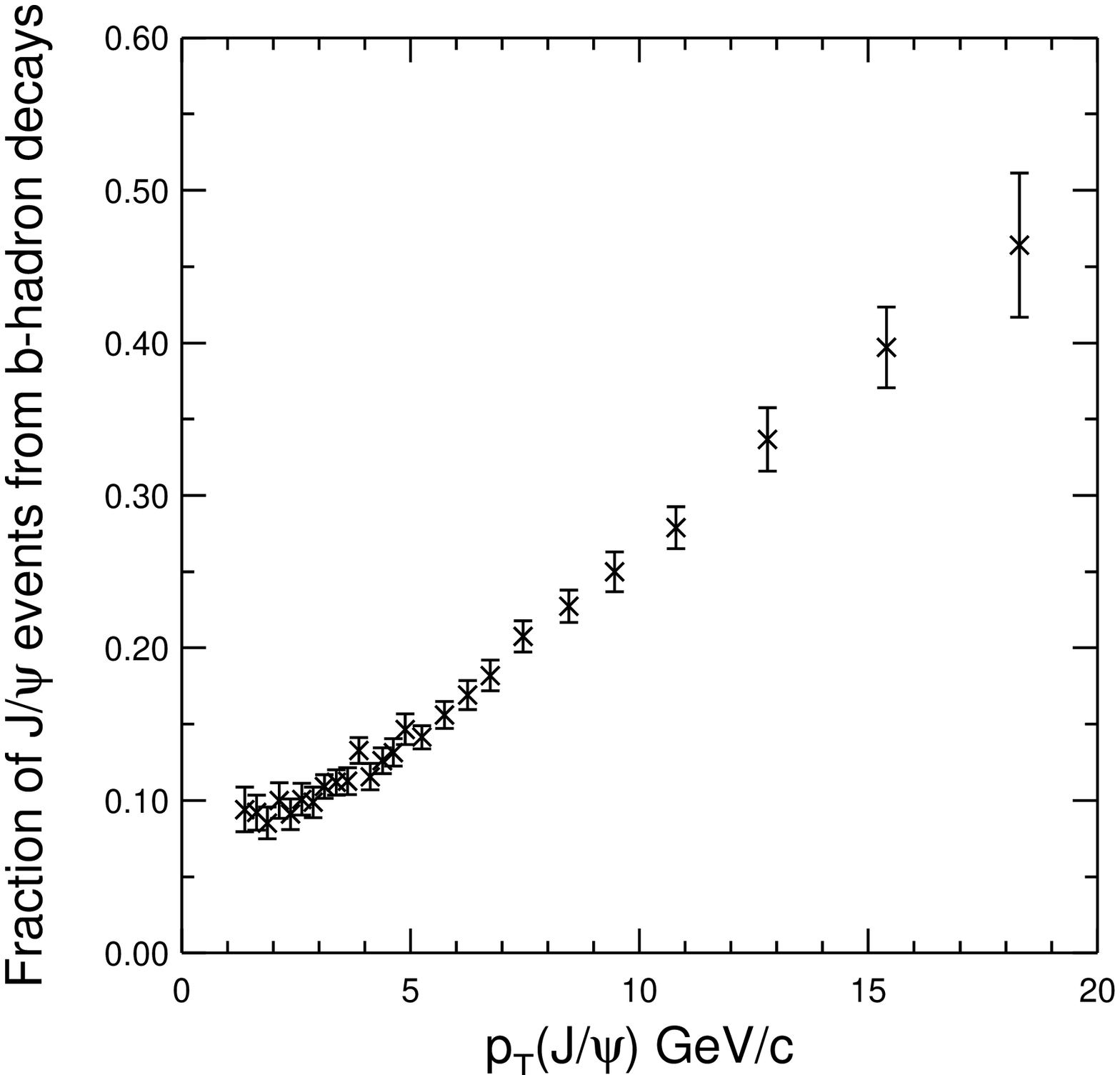,width=0.5\textwidth}}
\caption{Fraction of $J/\psi$ from $b$-hadron decays in the inclusive 
$J/\psi$ events of Run-II data as a function of $J/\psi$ transverse momentum. 
Error bars  include both statistical and systematic errors.   }
\label{fig_bfracbless}
\end{figure}
The fits to the $J/\psi$ pseudo-proper decay time in three sample $p_T$ ranges
are shown in Fig.~\ref{fig_bfracfita}, Fig.~\ref{fig_bfracfitb} and
Fig.~\ref{fig_bfracfitc}. These data correspond to a subset of the data in
the mass plots shown in Fig.~\ref{fig_mass1a} to Fig.~\ref{fig_mass1c}
which satisfy the SVX II tracking requirements.  
The values of the $b$-fractions from the fits with statistical and systematic
uncertainties  for  events with $J/\psi$ transverse
 momenta of  1.25\,GeV/$c$  to 20.0\,GeV/$c$ are listed 
in Table~\ref{tab_accsystb},  and the
distribution is shown in Fig.~\ref{fig_bfracbless}. 
This measurement of the $b$-fraction is used in
Section~\ref{sec_bxsec}, in conjunction with the measurement of the
inclusive $J/\psi$ cross section, to calculate the inclusive
$b$-hadron cross section. 

\begingroup
\squeezetable
\begin{table*}
\caption{
The fraction of $J/\psi$ events from decays of $b$-hadrons
and the corresponding acceptance. The first uncertainty on 
the $b$-fraction is the statistical uncertainty from the unbinned likelihood
fit and the second uncertainty is the combined systematic uncertainties
on the measurement of the $b$-fraction. The uncertainty on the
acceptance is the combined statistical uncertainty from Monte Carlo
statistics and the systematic uncertainty on the acceptance measurement. }
\begin{center}
\begin{tabular}{ccc} 
\hline \hline
$p_T(J/\psi)$ & Fraction from  & Acceptance  \\
 \,GeV/$c$       & $b$-hadrons    & $H_b\rightarrow J/\psi X$ \\ \hline    
$1.25-1.5$  &  $0.094 \pm 0.010 \pm 0.012$&
$0.01579\pm 0.00037$\\
$1.5-1.75$  &  $0.092 \pm 0.006 \pm 0.010$& $0.01981\pm 0.00029$\\
$1.75-2.0$  &  $0.085 \pm 0.006 \pm 0.009$& $0.02433\pm 0.00034$\\
$2.0-2.25$  &  $0.100 \pm 0.005 \pm 0.011$& $0.02842\pm 0.00032$\\
$2.25-2.5$  &  $0.091 \pm 0.005 \pm 0.010$& $0.03335\pm 0.00038$\\
$2.5-2.75$  &  $0.101 \pm 0.005 \pm 0.009$& $0.03864\pm 0.00059$\\
$2.75-3.0$  &  $0.099 \pm 0.005 \pm 0.008$& $0.04376\pm 0.00072$\\
$3.0-3.25$  &  $0.109 \pm 0.005 \pm 0.007$& $0.04940\pm 0.00081$\\
$3.25-3.5$  &  $0.112 \pm 0.005 \pm 0.008$& $0.05619\pm 0.00093$\\
$3.5-3.75$  &  $0.113 \pm 0.005 \pm 0.007$& $0.0611 \pm 0.0010$\\
$3.75-4.0$  &  $0.133 \pm 0.005 \pm 0.007$& $0.0666 \pm 0.0016$\\
$4.0-4.25$  &  $0.116 \pm 0.005 \pm 0.007$& $0.0736 \pm 0.0018$\\
$4.25-4.5$  &  $0.126 \pm 0.006 \pm 0.007$& $0.0815 \pm 0.0020$\\
$4.5-4.75$  &  $0.131 \pm 0.006 \pm 0.007$& $0.0891 \pm 0.0022$\\
$4.75-5.0$  &  $0.147 \pm 0.007 \pm 0.008$& $0.0960 \pm 0.0024$\\
$5.0-5.5$   &  $0.141 \pm 0.005 \pm 0.006$& $0.1065 \pm 0.0025$\\
$5.5-6.0$   &  $0.156 \pm 0.006 \pm 0.007$& $0.1198 \pm 0.0029 $\\
$6.0-6.5$   &  $0.169 \pm 0.007 \pm 0.007$& $0.1330 \pm 0.0032 $\\
$6.5-7.0$   &  $0.182 \pm 0.007 \pm 0.008$& $0.1476 \pm 0.0037 $\\
$7.0-8.0$   &  $0.208 \pm 0.006 \pm 0.009$& $0.1647 \pm 0.0055 $\\
$8.0-9.0$   &  $0.227 \pm 0.009 \pm 0.007$& $0.1813 \pm 0.0062 $\\
$9.0-10.0$  &  $0.250 \pm 0.011 \pm 0.008$& $0.1893 \pm 0.0068 $\\
$10.0-12.0$ &  $0.279 \pm 0.012 \pm 0.008$& $0.2022 \pm 0.0064 $\\
$12.0-14.0$ &  $0.337 \pm 0.019 \pm 0.009$& $0.2247 \pm 0.0072 $\\
$14.0-17.0$ &  $0.397 \pm 0.025 \pm 0.009$& $0.2462 \pm 0.011  $\\
$17.0-20.0$ &  $0.464 \pm 0.045^{+0.017}_{-0.011}$& $0.2538 \pm 0.0093 $\\	 
\hline \hline
\end{tabular}
\end{center}
\label{tab_accsystb}
\end{table*}
\endgroup

The uncertainties on the $b$-fractions are summarized in 
Table~\ref{tab_systb}. In the table, percentage errors on the absolute 
value of $b$-fraction are listed. 
Now we discuss the estimation of systematic 
uncertainties on  the $b$-fraction in detail. 
\begin{table}[!ht]
\caption{Sources of systematic uncertainties on the measurement of
the $b$-hadron fraction in inclusive $J/\psi$ decays as percentages of 
the absolute value.   In general, the $p_T$ dependent
uncertainties increase with decreasing $p_T$.}

\begin{center}
\begin{tabular}{lc} 
\hline \hline
Source & Systematic uncertainty \\ \hline
Resolution function model     &	$\pm (0.5-8)\%$ \\
Background function model     & $\pm (0-2)\%$   \\
Fit bias                      & $\pm (0-2)\%$   \\
MC production spectrum        & $\pm (2-7)\%$    \\
MC decay spectrum             & $\pm (0.5-3)\%$  \\ 
MC inclusive $H_b$ lifetime   & $\pm (0.5-4)\%$  \\ \hline
Total                         & $\pm (3-13)\%$  \\ \hline \hline
\end{tabular}
\end{center}
\label{tab_systb}
\end{table}

We have performed various tests to assess the accuracy of the
likelihood procedure. The fit shapes for signal and background are
histogramed into bins and compared to the binned data distributions. A
Kolmogorov-Smirnov (K-S) test~\cite{KS} is used to compare the fit and data
distributions to estimate the quality of the fit. The distribution of K-S
probability values for each fit in the different transverse momentum
ranges is  compared to the K-S probability distributions in a sample of
Monte Carlo experiments.  The distributions are found to be
consistent.  In addition to the K-S tests, the normalized residual, 
defined as the difference between the data and fit projection in the unit of 
one standard deviation of statistical error,   
is compared in every transverse momentum
range. Firstly, the data and fit projections are histogrammed using
an unequal pseudo-proper decay time bin size so that the number of data 
events in each bin is more than $20$ events to reduce statistical fluctuation.
Secondly, the normalized residual distributions are examined.  
The means and widths of the distributions in all transverse momentum ranges 
are examined. We find no obvious discrepancies between the fit projection
and data distributions.

Monte Carlo samples are also used to determine the potential bias on 
the $b$-fraction from the fitting procedure.   
The pseudo-proper decay time distributions and the invariant mass 
distributions from  signal and background  are  used to  
generate a set of 500 statistically independent
samples for each of the four $p_T$ bins of 1.25-1.5\,GeV/$c$, 
2.0-2.25\,GeV/$c$, 5.0-5.5\,GeV/$c$ and 10.0-12.0\,GeV/$c$.   
Five different values of the $b$-fraction, 5$\%$ to $13\%$,  are assumed 
and the number of events in each $p_T$ bin is chosen to match the data.
The fitted values of $b$-fractions are  found to agree with 
the generated values within 2$\%$ over the 
whole $p_T$ bins.  
Thus the systematic uncertainties   on  the 
$b$-fraction measurements due to fit bias are found to be less than 2$\%$.

The resolution function for the pseudo-proper decay time, 
$R(x'-x,s\sigma)$,  is modeled by a double Gaussian function 
where the dominant Gaussian width  is allowed to float and is 
determined by the fit to the  data in each  $p_T(J/\psi)$ bin.   
Other  parameters in the function  are fixed to 
the values obtained  from a binned fit to $L_{\rm xy}/\sigma(L_{\rm xy})$
averaged over all $p_T(J/\psi)$. The double Gaussian resolution
function is not an exact description of the resolution function shape
but only an approximate parameterization of many different resolution
effects.  Therefore, to estimate the systematic uncertainty due to the
resolution function  modeling, the maximum range of values for the 
ratios of  areas and widths of the two Gaussians  supported by the data 
are estimated. We find that the  ratios of the  second Gaussian 
to the dominant Gaussian  vary from $1.5$ to $2.5$ in width 
and 0.05 to 0.15 in area.  
The systematic uncertainty on the $b$-fraction from this source 
is largest in the lowest momentum bin, where the percentage
error is as large as  8$\%$,  
and decreases with increasing $p_T(J/\psi)$.

In the $J/\psi$ pseudo-proper decay time signal region, events are observed in
the distribution at long positive and negative lifetimes that are not
well described by the double Gaussian description of the resolution
function. The source of these long lived ``tails'' is unknown.  To
estimate the systematic uncertainty on the long lived tails not
modeled by the prompt signal double Gaussian, a box shaped function is 
added to the prompt $J/\psi$ $x$ signal distribution in the range  -2500
to 2500~$\mu$m. The height of the box is fixed in the fit using the number
of events in the data that are  in excess of the fit projection. The
$b$-fraction values returned from the fit with the box function are
used to estimate the systematic uncertainty from the tails that are
not modeled properly. We find the $b$-fraction values decrease by
about 5$\%$ in the lowest momentum bins when the box shape is added to
the prompt $J/\psi$ distribution. The excess modeled by the box can
also be assigned to the $b$-hadron signal which causes a systematic
increase. The change in the $b$-fraction decreases in the higher
transverse momentum bins.  

The fit was repeated with the background shape changed such that only
a positive and negative exponential is used with no symmetric
exponential. The differences in the $b$-fractions observed are
negligible.  The background parameters are extracted from a fit to the
sideband data distributions only, where the sidebands are chosen such
that no significant contribution is expected from the radiative
$J/\psi$ tail. The fit is repeated in each bin with the values of the
background parameters fixed to the values obtained from the sideband
fit. No significant difference between the value of the parameters
extracted is observed.  The difference in the $b$-fraction extracted
using the parameters obtained from the sideband fit is taken as a
systematic uncertainty. In the lowest and highest momentum bin the
percentage difference on $b$-fraction value extracted 
is $2$-$3\%$. The differences are less than $1\%$ in all other bins.

To study the dependence of the $b$-fractions  on the modeling
of the $b$-hadron spectrum used in the Monte Carlo, a flat
distribution in $p_T$ and $y$ of the $b$ production spectrum is used
to regenerate the $x$ distributions and the fits were repeated.  The
differences in the value of the $b$-fractions extracted from the
direct fit to Monte Carlo templates of $x$ produced from an input spectrum 
that is uniform in  $p_T(b)$ and  $y(b)$ are examined. The variation in the
$b$-fractions extracted in the range $1.25$ to $2.0$\,GeV/$c$ are the
largest, the maximum variation being an increase of 18$\%$ in the bin
$1.5$ to $1.75$\,GeV/$c$. The uniform  input spectrum is
unrealistic, therefore the systematic uncertainty is taken as one-half of
the size of the variation observed in the $b$-fraction when the flat
model is used. We assign systematic uncertainties of $7\%$, $3\%$
and $2\%$ for measurements in the transverse momentum ranges of 
1.25-3.0\,GeV/$c$, 3.0-8.0\,GeV/$c$ and 8.0-20.0\,GeV/$c$ respectively.

In addition, we examine the change in the $b$-fraction extracted when
varying the $H_b\rightarrow J/\psi X$ decay momentum spectrum 
while keeping the $H_b$ production momentum spectrum fixed. Two 
decay  spectrums, $H_b\rightarrow J/\psi({\rm direct}) X $ and 
$H_b\rightarrow J/\psi({\rm inclusive}) X$~\cite{CLEOPSI}, are used for this. 
The  percentage difference is found to be 2-4$\%$ in the
lowest momentum bins and $< 1\%$ for $p_T(J/\psi)>2.5$\,GeV/c.

The mix of hadrons and their respective lifetimes is a contributing
factor to the shape of the $J/\psi$ pseudo-proper decay time distributions. To
assess the systematic uncertainty due to the uncertainty on the
$b$-hadron average lifetime,  we vary the average lifetime in the Monte
Carlo by $11~\mu$m which is the size of the systematic uncertainty on
the average $b$-hadron lifetime measured at CDF in Run II. We find that
the measured $b$-fraction decreases in all transverse momentum bins
when the lifetime is increased. The fractional decrease is $4\%$ in
the lowest momentum bin and less than 
$1\%$ for bins with $p_T(J/\psi) > 12$\,GeV/$c$. 
The variation in the $b$-fraction observed when the 
average $b$-hadron lifetime is varied by $\pm 11 \ \mu$m is taken as a 
systematic uncertainty on the $b$-fraction measurement. 
Table~\ref{tab_systb} summarizes the sources of systematic uncertainties on 
the measurement of the $b$-hadron fraction as percentages of the absolute 
values.

\subsection{\boldmath Measurement of the Inclusive $b$-hadron
Cross Section ~\label{sec_bxsec}}
Since $J/\psi$ mesons from decays of bottom hadrons have a
different average spin alignment than an inclusive sample of $J/\psi$
mesons, we need to apply an acceptance correction to account for this
difference. In previous CDF measurements, the effective value of the
spin alignment parameter $\alpha_{\rm eff}$ of $J/\psi$ from $b$-hadron
decays was measured to be $\alpha_{\rm eff}(p_T(J/\psi) > 4.0 \ {\rm
\,GeV/}c ) = -0.09 \pm 0.10$~\cite{Jpsi_pol}, where $\alpha_{\rm eff}$ is
obtained by fitting $ \cos \theta_{\rm J/\psi}$, the angle between the
muon direction in the $J/\psi$ rest frame and the $J/\psi$ direction
in the lab frame, to the functional form $1+ \alpha_{\rm eff} \cdot \cos^2
\theta_{\rm J/\psi}$. More recent measurement on the spin alignment was 
done using $B\rightarrow J/\psi X$ events collected at the $\Upsilon(4S)$
resonance. The BaBar experiment measured
$\alpha_B=-0.196\pm0.044$ for $ p^*<1.1$\,GeV/$c$
and $\alpha_B=-0.592\pm0.032$ for $p^*>1.1$\,GeV/$c$~\cite{Babar}.
Here the decay angle of the $J/\psi$ is measured in the 
$\Upsilon(4S)$ rest frame and $p^*$ is the total $J/\psi$ momentum 
measured in the $\Upsilon(4S)$ rest frame. 

We opt to use the more precise  
result from the BaBar experiment in the acceptance calculations for 
$H_b \rightarrow J/\psi X$ events assuming it is applicable to 
the CDF environment where $b$-hadrons are produced in fragmentation 
with a large momentum range instead of produced at a fixed momentum 
as in $\Upsilon$ decays~\cite{Krey}. 
First, Monte Carlo events  are generated to have  the  $J/\psi$ helicity 
angle distributions in the $b$-hadron rest frame predicted from 
$\alpha_B$ values according to their  $p^*$ values.  Then,  values of the 
spin alignment parameter  $\alpha_{\rm eff}$ for events in 
each $J/\psi$ $p_T$ bin 
are obtained from fitting the $ \cos \theta_{\rm J/\psi}$ distributions of
these Monte Carlo events.  
The systematic errors on   $\alpha_{\rm eff}$ are obtained by 
varying the input values of  $\alpha_B$  in the process
according to measurement errors.  
This process gives a result consistent  with previous CDF measurement, 
albeit with smaller uncertainties. 
For example, a new and more precise value of 
 $\alpha_{\rm eff}  = -0.13 \pm 0.01$ for the $J/\psi$ events with 
$p_T(J/\psi)> 4.0 \ {\rm\,GeV/}c$ is obtained from this process.
Finally,  the acceptance  values, as listed in 
Table~\ref{tab_accsystb}, are calculated from the Monte Carlo
events generated with the derived spin alignment parameters 
in each $J/\psi$ $p_T$ bin. 

\begin{figure}[!h]
\centerline{\psfig{figure=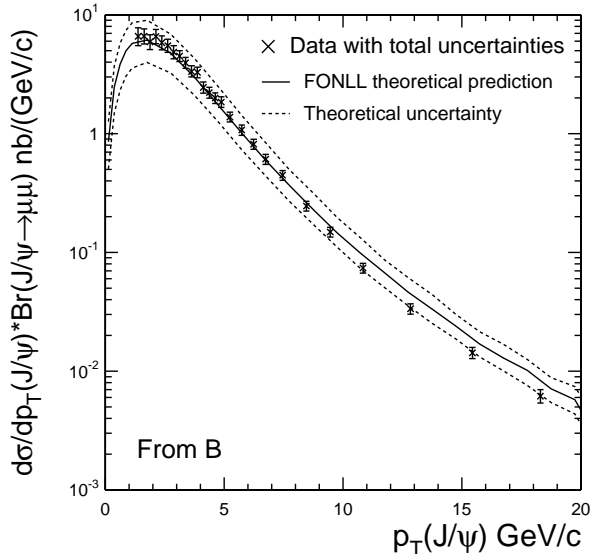,width=0.5\textwidth}}
\caption{Differential cross-section distribution of $J/\psi$ events
from the decays of $b$-hadrons as a function of $J/\psi$ transverse
momentum integrated over the rapidity range $|y|<0.6$. The crosses
with error bars are the data with systematic and statistical
uncertainties added including correlated uncertainties. The solid line
is the central theoretical values using the FONLL calculations
outlined in~\cite{MLM}, the dashed line is the theoretical
uncertainty.}
\label{fig_bxsec}
\end{figure}

The differential $b$-hadron cross sections are calculated in a similar 
way as that for the inclusive $J/\psi$. 
The $J/\psi$ yields in each $p_T$  bin listed 
in Table~\ref{tab_xsecsum} are multiplied with the  
$b$-fractions to obtain the corresponding $H_b \rightarrow J/\psi$ yields. 
The new acceptance values listed  in Table~\ref{tab_accsystb} are 
used while the $J/\psi$ reconstruction efficiencies and luminosity value 
stay the same. Most of the systematic uncertainties in the inclusive $J/\psi$ 
cross-section calculation carry over here without change except for those from 
the $J/\psi$ spin alignment on the acceptance which are 
estimated using errors on $\alpha_{\rm eff}$. In addition, the 
uncertainties from the $b$-fractions are also included in the 
systematic errors. 
The $J/\psi$ from $b$-hadron inclusive cross-section results with
statistical and systematic uncertainties are shown in
Table~\ref{tab_bxsec}. The differential cross section with all
statistical and systematic errors added is plotted 
in Fig.~\ref{fig_bxsec}. A recent QCD theoretical calculation using a fixed
order (FO) calculation with resummation of next-to-leading 
logs (NLL)~\cite{MLM} is overlaid. We discuss further the comparison with
theoretical calculations in Section~\ref{sec_discuss}.

\begingroup
\squeezetable
\begin{table*}
\caption{
The inclusive $H_b \rightarrow J/\psi X$ and prompt $J/\psi$ differential
cross sections as a function of transverse momentum of the $J/\psi$
with statistical and $p_T$ dependent systematic uncertainties. The
cross section in each $p_T$ bin is integrated over the rapidity range
$|y(J/\psi)| < 0.6$ The fully correlated systematic uncertainty,
$syst_{fc}=6.9\%$, from the measurement of the inclusive $J/\psi$ cross
section needs to combined with the $p_T$ dependent systematic uncertainties.}
\begin{center}
\begin{tabular}{cccc}
\hline \hline
$p_T(J/\psi)$ & $\langle p_T(J/\psi)\rangle$  & $\frac{d\sigma}{dp_T} \cdot Br$ (nb/\,GeV/$c$) & $\frac{d\sigma}{dp_T} \cdot Br$ (nb/\,GeV/$c$)  \\  
(\,GeV/$c$) & (\,GeV/$c$) &  $J/\psi$ from $b$ & Prompt $J/\psi$\\ \hline
$1.25-1.5$  & $ 1.38 $& $  6.60 \pm 0.70(stat)^{+0.77}_{-0.67} (syst_{p_T})$ & $  66.8 \pm 1.5(stat)^{+9.2}_{-9.1} (syst_{p_T})$\\
$1.5-1.75$  & $ 1.63 $& $  6.62 \pm 0.44^{+0.71}_{-0.62} $ & $ 68.6 \pm 1.5^{+8.2}_{-8.0} $\\	
$1.75-2.0$  & $ 1.87 $& $  5.93 \pm 0.38^{+0.62}_{-0.56} $ & $ 67.0 \pm 1.3^{+7.9}_{-7.7} $\\	
$2.0-2.25$  & $ 2.13 $& $  6.58 \pm 0.34^{+0.67}_{-0.56} $ & $ 62.5 \pm 0.7^{+7.5}_{-7.4} $\\	
$2.25-2.5$  & $ 2.38 $& $  5.83 \pm 0.30^{+0.57}_{-0.50} $ & $ 61.5 \pm 0.9^{+7.3}_{-7.2} $\\	
$2.5-2.75$  & $ 2.62 $& $  5.50 \pm 0.26^{+0.51}_{-0.45} $ & $ 52.1 \pm 0.8 \pm {5.2} $\\	
$2.75-3.0$  & $ 2.87 $& $  4.86 \pm 0.23^{+0.44}_{-0.38} $ & $ 47.1 \pm 0.7 \pm {4.4} $\\	
$3.0-3.25$  & $ 3.12 $& $  4.50 \pm 0.20^{+0.25}_{-0.21} $ & $ 39.1 \pm 0.6 \pm {3.0} $\\	
$3.25-3.5$  & $ 3.38 $& $  3.94 \pm 0.17^{+0.23}_{-0.18} $ & $ 33.4 \pm 0.5 \pm 2.8 $\\	
$3.5-3.75$  & $ 3.62 $& $  3.34 \pm 0.15^{+0.21}_{-0.16} $ & $ 28.2 \pm 0.4 \pm 2.1 $\\	
$3.75-4.0$  & $ 3.87 $& $  3.28 \pm 0.14 \pm 0.16 $ & $  22.9 \pm 0.3 \pm 1.6 $\\	
$4.0-4.25$  & $ 4.12 $& $  2.45 \pm 0.11 \pm 0.15 $ & $  20.1 \pm 0.4 \pm 1.5$\\	
$4.25-4.5$  & $ 4.38 $& $  2.22 \pm 0.10 \pm 0.11 $ & $  16.5 \pm 0.3 \pm 1.2 $\\	
$4.5-4.75$  & $ 4.62 $& $  1.99 \pm 0.09 \pm 0.10 $ & $  14.1 \pm 0.3 \pm 1.0 $\\	
$4.75-5.0$  & $ 4.88 $& $  1.84 \pm 0.08 \pm 0.10 $ & $  11.5 \pm 0.3\pm 0.8 $\\	
$5.0-5.5$   & $ 5.24 $& $  1.38 \pm0.05  \pm 0.06 $ & $  8.92 \pm 0.13\pm 0.52 $\\	
$5.5-6.0$   & $ 5.74 $& $  1.07 \pm0.04  \pm 0.05 $ & $  6.21 \pm 0.10\pm 0.37 $\\	
$6.0-6.5$   & $ 6.24 $& $ 0.817 \pm0.031 \pm 0.038$ & $  4.29 \pm 0.07\pm 0.24 $\\	
$6.5-7.0$   & $ 6.74 $& $ 0.610 \pm0.025 \pm 0.026$ & $  2.93 \pm 0.06\pm 0.17 $\\	
$7.0-8.0$   & $ 7.45 $& $ 0.447 \pm0.014 \pm 0.022$ & $  1.82 \pm 0.02\pm 0.11 $\\	
$8.0-9.0$   & $ 8.46 $& $ 0.246 \pm0.009 \pm 0.010$ & $  0.894 \pm 0.015\pm 0.047 $\\	
$9.0-10.0$  & $ 9.46 $& $ 0.149 \pm0.007 \pm 0.006$ & $  0.473 \pm 0.010\pm 0.024 $\\	
$10.0-12.0$ & $ 10.8 $& $ 0.074 \pm0.003 \pm 0.003$ & $  0.204 \pm 0.004\pm 0.010 $\\	
$12.0-14.0$ & $ 12.8 $& $ 0.034 \pm0.002 \pm 0.001$ & $  0.069 \pm 0.002\pm 0.003 $\\	
$14.0-17.0$ & $ 15.4 $& $ 0.0143\pm0.0009 \pm 0.0007 $ & $  0.023 \pm 0.001\pm 0.001 $\\ 
$17.0-20.0$ & $ 18.3 $& $ 0.0062\pm0.0006 \pm 0.0004 $ & $  0.0078 \pm 0.0006\pm 0.0006 $\\     
\hline \hline
\end{tabular}
\end{center}
\label{tab_bxsec}
\end{table*}
\endgroup

An integration of the differential $b$-hadron cross-section results in
Table~\ref{tab_bxsec} gives the total cross section
\begin{eqnarray}
\nonumber
 \sigma(p\bar{p}\rightarrow H_b X, p_T(J/\psi) > 1.25~{\rm GeV}/c,
|y(J/\psi)|<0.6) \\ \nonumber
\cdot Br(H_b \rightarrow J/\psi X) \cdot Br(J/\psi
\rightarrow \mu \mu) \\ 
=  \BXSECBR \ {\rm nb}. 
\end{eqnarray}
The systematic uncertainty
quoted includes the fully correlated uncertainty of $6.9\%$ obtained
from the inclusive $J/\psi$ cross-section measurement. We correct 
the integrated cross section extracted above 
for the branching fraction 
$Br(J/\psi \rightarrow \mu \mu) = 5.88 \pm
0.10\%$~\cite{PDG} to obtain
\begin{eqnarray}
\nonumber
 \sigma(p\bar{p}\rightarrow H_b, ~H_b \rightarrow J/\psi,  
p_T(J/\psi) > 1.25 {\rm \ \,GeV}/c,|y(J/\psi)|<0.6)  \\ 
=  \BXSEC \ \mu{\rm b}. 
\end{eqnarray}

\begin{figure}[!h]
\centerline{\psfig{figure=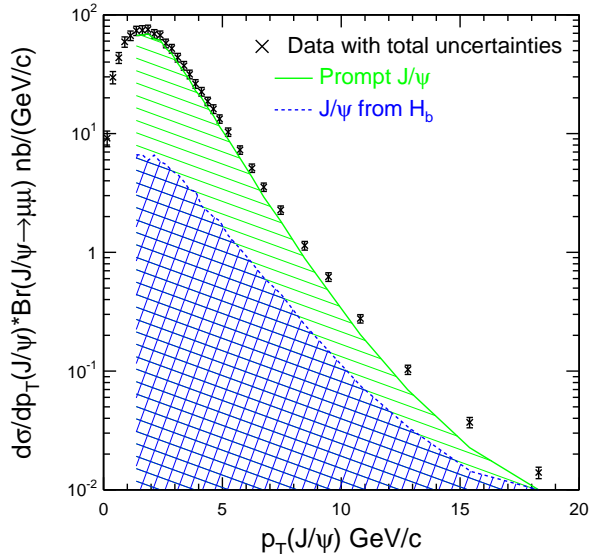,width=0.5\textwidth}
}
\caption{The inclusive $J/\psi$ cross section as a function of $J/\psi$
$p_T$ integrated over the rapidity range $|y|<0.6$ is plotted as
points with error bars where all uncertainties have been added.
The hatched histogram indicates the contribution to the cross section
from prompt charmonium production. The cross-hatched histogram is the
contribution from decays of $b$-hadrons.}
\label{fig_xsecall}
\end{figure}

We also extract the prompt $J/\psi$ cross section by subtracting the
cross section of $H_b \rightarrow J/\psi X$ from the inclusive $J/\psi$
cross section. This calculation is applied to all $J/\psi$ with
$p_T>1.25$\,GeV/c where we are able to extract the $b$-fraction. The
results are shown in Table~\ref{tab_bxsec} and in
Fig.~\ref{fig_xsecall}. The systematic uncertainties on the prompt
$J/\psi$ cross section are taken to be the uncertainties on the
inclusive cross section added in quadrature with the uncertainties on the
measured $b$-fractions.
We find the integrated cross section of prompt $J/\psi$ to be:
\begin{eqnarray}
\nonumber
 \sigma (p\bar{p}\rightarrow J/\psi_{p} X, p_T(J/\psi) > 1.25 {\rm \ \,GeV}/c,
|y(J/\psi)|<0.6)  \\ 
=  \PROMPTXSEC \ \mu{\rm b}, 
\end{eqnarray}
where $J/\psi_{p}$ denotes a prompt $J/\psi$ and where we have
corrected for the $J/\psi \rightarrow \mu \mu$ branching fraction.

The differential $b$-hadron cross section  as a function of $p_T(H_b)$
is extracted from the measured differential cross sections of $H_b
\rightarrow J/\psi X$ by utilizing the decay kinematics of charmonium
produced in $b$-hadron decays. 

The procedure starts with the calculation of contributions to the
cross section of $b$-hadrons in a given $p_T(H_b)$ bin from $J/\psi$
events in the range  $1.25 <p_T(J/\psi)<20$\,GeV/$c$, where we measured
the fractions of $J/\psi$ mesons from $b$ decays.
Since $b$-hadrons with as little as zero momenta  
produce $J/\psi$ mesons with momenta  as large 
as 2\,GeV/$c$,  the measured cross section  in  this $p_T(J/\psi)$ range  
is sensitive to the  complete $p_T (H_b)$ spectrum.
The total contribution to the cross section in
the $i^{th}$ bin in $p_T(H_b)$ from events in the accessible
$p_T(J/\psi)$ range, labeled as the raw cross section $\sigma_i (raw)$, is
given by
\begin{equation}
\sigma_i (raw) = \sum_{\rm j=1}^{N} w_{ij} \sigma_j (J/\psi),
\end{equation}
where $\sigma_j (J/\psi)$ is the cross section of 
$J/\psi$ mesons from $H_b$ in the $j^{th}$ $p_T(J/\psi)$ bin and $w_{ij}$ is 
the fraction of $H_b$ events in the $i^{th}$ $p_T(H_b)$ bin relative to the 
total in the $j^{th}$ $p_T(J/\psi)$ bin. 
The sum of the weights $w_{ij}$ in each $p_T(J/\psi)$ bin is 
normalized to 1. 
The raw cross section is  corrected for the acceptance due to the 
limited  $J/\psi$ $p_T$  range  to obtain 
the differential $b$-hadron cross section, $\sigma_i(H_b)$, in 
the $i^{th}$ $p_T(H_b)$ bin,
\begin{equation}
\sigma_i (H_b) = \frac{\sigma_i (raw)}{ f_{\rm \sigma}^i } 
=\frac{\sum_{\rm j=1}^{N} w_{\rm ij} \sigma_j (J/\psi) }
{f_{\rm \sigma}^i},
\end{equation}
where $f_{\rm \sigma}^i$ is the fraction of bottom hadrons in 
the $i^{th}$  $p_T(H_b)$ bin that give rise to a $J/\psi$ with a transverse
momentum in the range 1.25 to 20\,GeV/$c$ and rapidity in the range
$|y(J/\psi)| < 0.6$.  Monte Carlo simulations are used to calculate
the weighting factors, $w_{ij}$, and acceptance correction
factors, $f_{\rm \sigma}^i$.  In the simulation, the decay
spectrum of $H_b\rightarrow J/\psi X$ obtained from 
references~\cite{CLEOPSI} and \cite{Babar} is used.  
The calculation is repeated in an iteration process: 
at each pass the input production spectrum used in the
Monte Carlo is the spectrum measured in the previous iteration and 
a  $\chi^2$ comparison is made between the input and output spectrums. 
The process terminates when the $\chi^2$ comparison reach the precision limit. 
This procedure is found to be insensitive to the initial production
spectrum shape.

The statistical uncertainty in each  $p_T(H_b)$ bin is given by:
\begin{equation}
\delta_{\rm stat}(\sigma_i(H_b)) = \frac{1}{f_{\rm \sigma}^i}
\sqrt{ \sum_{j=1}^{N} w_{ij} 
\delta_{\rm stat}^2(\sigma_j(J/\psi))}.
\end{equation}
The systematic uncertainties are taken as just the simple weighted
sum of the systematic errors from the differential $H_b
\rightarrow J/\psi$ cross-sections measurements,
\begin{equation}
\delta_{\rm syst}(\sigma_i(H_b)) = \frac{1}{f_{\rm \sigma}^i}\sum_{j=1}^{N} w_{ij} \delta_{\rm syst}(\sigma_j(J/\psi)).
\end{equation}

\begin{figure}[!h]
\centerline{\psfig{figure=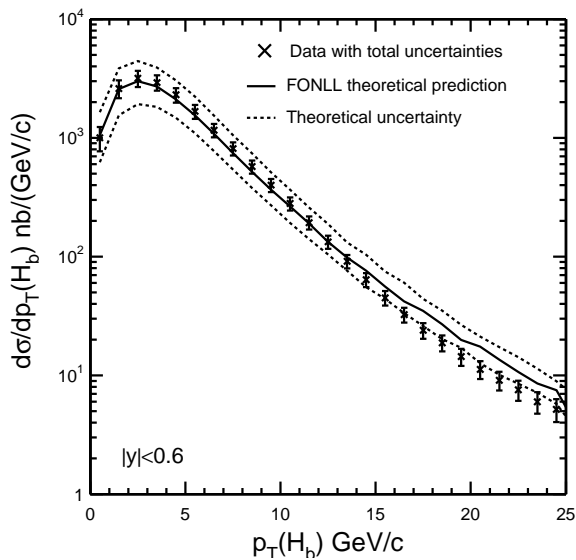,width=0.5\textwidth}}
\caption{Differential cross-section distribution of 
$b$-hadron production as a function of $b$-hadron transverse
momenta. The crosses with error bars are the data with systematic and
statistical uncertainties added, including correlated
uncertainties. The solid line is the central theoretical values using
the FONLL calculations outlined in~\cite{MLM}, the dashed line is the
theoretical uncertainty.}
\label{fig_bxsec5}
\end{figure}

The extracted differential cross section of $b$-hadrons over the
 transverse momentum range from 0 to 25\,GeV/$c$ is shown in
 Fig.~\ref{fig_bxsec5}. The cross section has been corrected for the
 branching fractions, $Br(H_b\rightarrow J/\psi X) = 1.16\pm 0.10 \%$
 and $Br(J/\psi \rightarrow \mu \mu) = 5.88 \pm 0.10 \%$~\cite{PDG}, and
 divided by two to obtain the single $b$-hadron differential cross
 section.  We integrate the differential cross section extracted above
 to obtain the single $b$-hadron inclusive cross section. We find
 the total inclusive single $b$-hadron cross section is
\begin{equation}
\sigma(p\bar{p}\rightarrow H_b X, |y|<0.6) = \TOTXSECB \ \mu{\rm b}.
\end{equation}

\section{Discussion \label{sec_discuss}}
We have measured the inclusive $J/\psi$ and $b$-hadron cross sections in
$p\bar{p}$ interactions at $\sqrt{s} = 1960$\,GeV in the central
rapidity region of $|y|<0.6$. For the first time, the cross section
has been measured over the full transverse momentum range (0-20
\,GeV/$c$).

For comparison to  Run~I measurements at 
$\sqrt{s}=1800$\,GeV ~\cite{RUNIPSI},  we consider the cross-section
measurements in the range $p_T (J/\psi)>5.0$\,GeV/$c$ and pseudo-rapidity
$|\eta(J/\psi)|< 0.6$. We measure the inclusive $J/\psi$ cross section
 at $\sqrt{s}=1960$\,GeV to be 
\begin{eqnarray}
\nonumber
\sigma(p \overline{p} \rightarrow
J/\psi X)_{1960} \cdot Br(J/\psi \rightarrow \mu \mu) && \\
 =  \XSECGTFIVEETA \ {\rm nb}. && 
\end{eqnarray}
The CDF Run I measurement at   $\sqrt{s}=1800$\,GeV      
was found to be 
\begin{eqnarray}
\nonumber
\sigma(p \overline{p} \rightarrow J/\psi X)_{1800} \cdot
Br(J/\psi \rightarrow \mu \mu) \\ 
= 17.4 \pm 0.1(stat)^{+2.6}_{\rm -2.8} (syst) \ {\rm nb}. 
\end{eqnarray}

\begin{figure}[!h]
\centerline{\psfig{figure=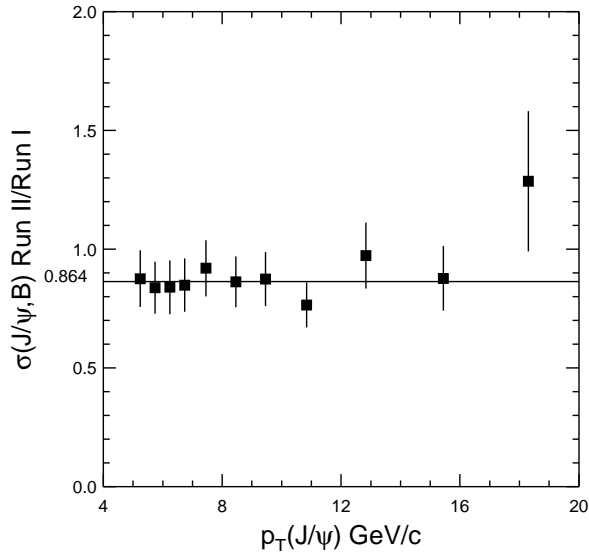,width=0.5\textwidth}}
\caption{Ratio of the differential cross-section distributions of $J/\psi$ events
from the decays of $b$-hadrons as a function of $J/\psi$ transverse
momentum from CDF Run I and Run II. }
\label{fig_comp1}
\end{figure}

We measure the  cross section of $J/\psi$ events from $H_b$ decays with
$p_T(J/\psi) > 5$\,GeV/$c$ and $|\eta(J/\psi)| < 0.6$ to be 
\begin{eqnarray}
\nonumber
\sigma(p \bar{p} \rightarrow H_b X)_{1960} \cdot Br(H_b
\rightarrow J/\psi X) \cdot Br(J/\psi \rightarrow \mu \mu) \\
=\BXSECBRFIVEETA  \ {\rm nb}.
\end{eqnarray}
The equivalent Run I measurement
~\cite{RUNIPSI} was found to be 
\begin{eqnarray}
\nonumber
\sigma(p \bar{p} \rightarrow H_b
X)_{1800}  \cdot Br(H_b
\rightarrow J/\psi X) \cdot Br(J/\psi \rightarrow \mu \mu) \\
= 3.23 \pm 0.05(stat) ^{+0.28}_{\rm -0.31}(syst) \ {\rm nb}. 
\end{eqnarray}
Although the Run II $J/\psi$ and $b$-hadron cross sections are
measured at a higher center-of-mass energy, and it is expected that
the production cross sections increase by approximately $10\%$, the
Run I and Run II measurements are consistent within measurement
uncertainties.  The ratio of the Run II to Run I differential
$b$-hadron cross-section measurements as a function of $p_T(J/\psi)$
is shown in Fig.~\ref{fig_comp1}. No difference in the shape of the
cross section is observed.

\begin{figure}[!h]
\centerline{\psfig{figure=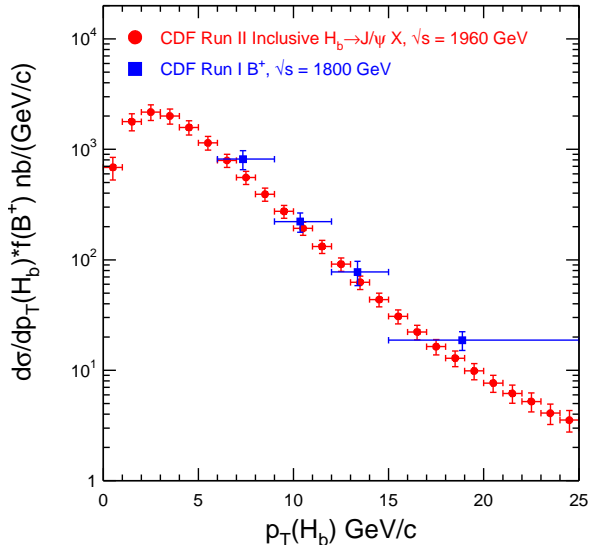,width=0.5\textwidth}}
\caption{The differential cross-section distributions of $b$-hadrons 
as a function of $b$-hadron momentum from the measurement of the $B^+$
meson cross section in CDF Run I ~\cite{CDFBXsec4} and $b$-hadron
inclusive cross section extracted in this analysis (Run II).  The
differential cross sections showed are integrated over the rapidity
range $|y(H_b)|< 1.0$.}
\label{fig_comp2}
\end{figure}

In Fig.~\ref{fig_comp2}, the $B^+$ differential cross section
previously measured by CDF at $\sqrt{s}=1800$\,GeV for $|y|<1.0$
~\cite{CDFBXsec4} is compared with our newer  measurement of the inclusive
$b$-hadron differential cross section at $\sqrt{s}=1960$\,GeV 
extracted from the measurement of the cross section of $J/\psi$ events
from $b$-hadron decays. For the purpose of this comparison, the CDF
Run II inclusive $b$-hadron cross section is multiplied by the
fragmentation fraction of $B^+$ mesons, where the result from LEP 
experiments is used~\cite{BFRAG}.
In addition, the Run II $b$-hadron inclusive cross
section is scaled up by a factor of 1.67 to extend the measurement
to $|y|<1.0$ where we have assumed the rapidity distribution is uniform
in the region $|y|<1.0$.
As shown in Fig.~\ref{fig_comp2}, we find good agreement between
the Run II extracted measurement of the $b$-hadron cross section and
the direct measurement of the $B^+$ cross section in Run I.

In Fig.~\ref{fig_bxsec} and Fig.~\ref{fig_bxsec5}, we compare our
measurement to a QCD calculation of the $b$-hadron cross section
by Cacciari {\it et al.}~\cite{MLM}. This 
calculation uses a fixed-order approach
with a next-to-leading-log resummation and a new technique to extract
the $b$-hadron fragmentation function from LEP data~\cite{BTHEORY4,MLM}. The 
single $b$-hadron cross section from this
FONLL calculation using the CTEQ6M parton distribution 
functions~\cite{CTEQ} is 
$\sigma^{FONLL}_{\rm (|y|<0.6)}=16.8^{+7.0}_{-5.0}~\mu{\rm b}$ which is in good agreement with our measurement of $\TOTXSECB \ \mu{\rm b}$.

We also compare this result to the QCD calculation described in
reference ~\cite{BTHEORY3}. This calculation employs a factorization
scheme where the mass of the quark is considered negligible and a
different treatment of the $b$-hadron fragmentation function is
used. The cross-section calculation in \cite{BTHEORY3} is repeated
using $\sqrt{s} = 1960$\,GeV/$c$ and the MRST2001 parton
distribution functions~\cite{MRST}. The central value of the
calculated cross section integrated over the rapidity range $|y|<0.6$
and $p_T(J/\psi)> 5.0$\,GeV/$c$ is $\sigma(p\bar{p}
\rightarrow H_b X,|y|<0.6) \cdot Br(H_b \rightarrow J/\psi X) \cdot
Br(J/\psi \rightarrow \mu \mu) = 3.2 \ {\rm nb}$~\cite{BERNDT}  which 
is in good agreement with our result  of $\BXSECBRFIVEGEV~{\rm nb}$

A more complete discussion of the changes in QCD calculations can be
found in references ~\cite{BTHEORY4, MLM, BTHEORY2}.  Updated
determinations of proton parton densities and bottom quark
fragmentation functions have brought the QCD calculations into better
agreement with the CDF measurements of the total $b$-hadron cross
section and the $b$-hadron $p_T$ distribution.


\section{Summary}
We have measured the inclusive central $J/\psi$ cross section in
$p\bar{p}$ interactions at $\sqrt{s} = 1960$\,GeV.  The cross section
has been measured over the full transverse momentum range.
for the first time.  We find the integrated inclusive $J/\psi$ cross
section in the central rapidity range to be 
\begin{eqnarray}
\nonumber
\sigma(p
\overline{p} \rightarrow J/\psi X,\mid y(J/\psi) \mid < 0.6) \\
= \TOTXSEC \ \mu{\rm b},
\end{eqnarray} 
after correcting for
$Br(J/\psi \rightarrow \mu \mu) = 5.88 \pm 0.10 \%$ ~\cite{PDG}.  

Using the long lifetime of $b$-hadrons to separate that portion of
the $J/\psi$ cross section that is from decays of $b$-hadrons,  we have
measured the cross section of $J/\psi$ mesons from $b$-hadron decays
for $J/\psi$ transverse momenta greater than $1.25$\,GeV/$c$. The
integrated $H_b \rightarrow J/\psi X$ cross section, including both
hadron and anti-hadron states, is 
\begin{eqnarray}
\nonumber
\sigma(p\bar{p}\rightarrow H_b, ~H_b \rightarrow J/\psi X, p_T(J/\psi) 
> 1.25 {\rm \ \,GeV}/c,
|y(J/\psi)|<0.6) \\ = \BXSEC \ \mu {\rm b},
\end{eqnarray}
after correcting for the
branching fraction
$Br(J/\psi \rightarrow \mu \mu) = 5.88 \pm 0.10\%$~\cite{PDG}.  

The measurement of the $J/\psi$ cross section from $b$-hadron decays
probes $b$-hadron transverse momenta down to zero.  We have extracted
the first measurement of the total central $b$-hadron cross section in
$p\bar{p}$ collisions from the measurement of the $b$-hadron cross
section with $J/\psi$ transverse momenta greater than 
$ 1.25$\,GeV/$c$ using Monte
Carlo models. We find the total single $b$-hadron
cross section integrated over all transverse momenta to be
\begin{equation}
\sigma(p\bar{p}\rightarrow H_b X, |y|<0.6) =
\TOTXSECB \ \mu{\rm b}.
\end{equation}




\begin{acknowledgments}
We thank the Fermilab staff and the technical staffs of the participating 
institutions for their vital contributions. This work was supported by 
the U.S. Department of Energy and National Science Foundation; the Italian 
Istituto Nazionale di Fisica Nucleare; the Ministry of Education, Culture, 
Sports, Science and Technology of Japan; the Natural Sciences and Engineering Research Council of Canada; the National Science Council of the Republic 
of China; the Swiss National Science Foundation; the A.P. Sloan 
Foundation; the Bundesministerium fuer Bildung und Forschung, Germany; the 
Korean Science and Engineering Foundation and the Korean Research
 Foundation; the Particle Physics and Astronomy Research Council and 
the Royal Society, UK; the Russian Foundation for Basic Research; the 
Commision Interministerial de Ciencia y Tecnologia, Spain; and in part 
by the European Community's Human Potential Programme under 
contract HPRN-CT-2002-00292, Probe for New Physics.

\end{acknowledgments}


\end{document}